\documentclass[modern]{aastex63}
\usepackage{amssymb, amsmath}
\usepackage{color}
\usepackage{graphicx}
\usepackage{morefloats}
\usepackage{hyperref}

\begin{document}

\title{The Red Supergiant Content of M31 and M33}

\author{Philip Massey}
\affiliation{Lowell Observatory, 1400 W Mars Hill Road, Flagstaff, AZ 86001}
\affiliation{Department of Astronomy and Planetary Science, Northern Arizona University, Flagstaff, AZ, 86011-6010}
\email{massey@lowell.edu}

\author{Kathryn F. Neugent}
\affiliation{Department of Astronomy, University of Washington, Seattle, WA, 98195}
\affiliation{Lowell Observatory, 1400 W Mars Hill Road, Flagstaff, AZ 86001}
\email{kneugent@uw.edu}

\author{Emily M. Levesque}
\affiliation{Department of Astronomy, University of Washington, Seattle, WA, 98195}
\email{emsque@uw.edu}

\author{Maria R. Drout}
\affiliation{David A. Dunlap Department of Astronomy and Astrophysics, University of Toronto, 50 St. George Street, Toronto, Ontario, M5S 3H4 Canada}
\affiliation{Observatories of the Carnegie Institution for Science, 813 Santa Barbara St., Pasadena, CA 91101, USA}
\email{maria.drout@utoronto.ca}

\author{St\'{e}phane Courteau}
\affiliation{Department of Physics, Engineering Physics \& Astronomy, Queen's University, Kingston, ON  K7L 3N6  Canada}
\email{courteau@queensu.ca}

\begin{abstract}
We identify red supergiants (RSGs) in our spiral neighbors M31 and
M33 using near-IR (NIR) photometry complete to a luminosity limit of $\log
L/L_\odot=4.0$. Our archival survey data cover 5 deg$^2$ of
M31, and 3 deg$^2$ for M33, and are likely spatially complete for
these massive stars.  {\it Gaia} is used to remove foreground stars, after
which the RSGs can be separated from asymptotic giant branch (AGB) stars
in the color-magnitude diagram. The photometry is used to derive
effective temperatures and bolometric luminosities via MARCS stellar
atmosphere models.  The resulting H-R diagrams show superb agreement
with the evolutionary tracks of the Geneva evolutionary
group. Our census includes 6400 RSGs in M31 and 2850 RSGs in M33 within
their Holmberg radii; by contrast, only a few hundred RSGs are known
so far in the Milky Way.  Our catalog serves as the basis for 
a study of the RSG binary frequency being published separately, as well
as future studies relating to the evolution of massive stars. Here we use the matches
between the NIR-selected RSGs and their optical counterparts to show that the
apparent similarity in the reddening of OB stars in M31 and M33 is
the result of Malmquist bias; the average extinction in M31 is likely
higher than that of M33.    As expected, the distribution
of RSGs follows that of the spiral arms, while the much older AGB
population is more uniformly spread across each galaxy's disk.
\end{abstract}

\section{Introduction}

Red supergiants (RSGs) are the evolved descendants of 8-30$M_\odot$ stars.   Once their main-sequence
progenitor OB stars exhaust the hydrogen in their cores, hydrogen fusion in a shell around the inert
helium core results in rapid expansion of their outer layers.  The stars zip across the H-R diagram, cooling and expanding tremendously.  By the time the stars have reached equilibrium at the Hayashi
limit, helium has ignited in the cores.  At this point the star has expanded to radii of 500-1800$R_\odot$, with
effective temperatures of 3500-4500~K. (For a recent exposition, the reader is referred to \citealt{LevesqueRSGs}; see also 
\citealt{Dorda2016} and \citealt{Sylvia}.)
Unevolved single stars of higher masses ($>\sim 30M_\odot$) do not evolve to the RSG phase, but instead
spend their He-burning years as Wolf-Rayet (WR) stars after experiencing extensive mass loss, possibly aided by first undergoing a luminous blue variable (LBV) phase.   Stars of lower mass ($<8M_\odot$) evolve instead to  red {\it giants}, (not  {\it supergiants}), and are not considered massive stars, as they are not expected to end as a core-collapse supernova.  Binary interactions may modify this picture, most notably providing a channel to the WR stage by stars that would otherwise become RSGs (see, e.g., \citealt{1996LIACo..33...55D,sana12,2017IAUS..329....3M}).


An accurate knowledge of the RSG content of nearby galaxies is useful for testing models of massive star evolution. As \citet{Kipp} put it, these evolved stages act a ``sort of magnifying glass" on stellar models, ``revealing relentlessly the faults of calculations of earlier phases."  
For instance, \citet{Maeder80} first pointed out that the relative number of RSGs and WRs should be a very sensitive function of metallicity, and indeed \citet{MasseyARAA} found such an effect observationally, but showed there was remarkably poor quantitative agreement with theoretical predictions.  However, in retrospect, neither the RSG content nor WR numbers were known accurately at that time.  Further complications in interpreting these numbers have been discussed by \cite{2018ApJ...867..125D} and \cite{2020MNRAS.497.2201S}. 

 Another example of how evolved massive stars can help us is in improving our understanding of binary influence on massive star evolution, and its metallicity dependence.  The RSG binary frequency can be compared to that of unevolved massive stars, providing a stringent test on the interaction rates.  About ten RSG binaries are known in the Milky Way \citep{NeugentRSGBinII}. A spectroscopic survey of cool supergiants in the Small and Large Magellanic Clouds (SMC, LMC)  by \citet{2015A&A...578A...3G} identified several likely physical binaries.  Directed searches for RSG binaries by \citet{NeugentRSGBinII}  led to the discoveries of  of 87 in LMC, M31, and M33.  Follow-up work in the LMC identified \citet{LMCBins} 37 more, and allowed a determination of the RSG binary frequency in the LMC.  In a paper being submitted contemporaneously with this one, \citet{M31M33RSGBins} has used the sample of RSGs identified here, along with new spectroscopy and {\it HST} UV data, to establish the RSG binary frequency in M31 and M33, demonstrating a significant metallicity dependence.
  
To support this work, we must first identify relatively complete populations of RSGs in galaxies of various metallicities. Work by \citet{Yang2019} has recently identified the RSG content in the Small Magellanic Cloud (SMC).  \citet{LMCBins} subsequently used a similar technique to obtain a census of the RSG population of the LMC. In this paper, we determine a luminosity-limited, spatially complete sample of RSGs for the nearby spiral galaxies M31 and M33.  Both galaxies have long been known to be rich in massive stars; the results of recent spectroscopic survey can be found in \citet{BigTable}.  

These galaxies provide a high-metallicity complement to studies in the SMC and LMC, where the young-population metallicities are 25\% and 50\% of the solar value \citep{Russell1990}.  For M31 the metallicity (at 10~kpc) is 1.5-2$\times$ solar with only a modest galactocentric gradient \citep{Zaritsky1994,Venn2000,Sanders}, while the metallicity of M33 varies from super-solar in the central regions to SMC-like in the outer regions \citep{Zaritsky1994,Urbaneja2005}; see further discussion in \citealt{NeugentM33} and \citealt{M31M33RSGBins}.
 
Studies of the RSG content of nearby galaxies have traditionally been hampered by two problems.  The first of these is foreground contamination.  Although such contamination is relatively low ($<$10\%) for RSGs in the Magellanic Clouds \citep{Massey2002,NeugentLMC}, it is 50\% or more for the more distant members of the Local Group, such as M31 and M33 \citep{MasseyRSGs}. Optical two-color diagrams can help distinguish between foreground dwarfs and extragalactic supergiants \citep{MasseyRSGs}, although radial velocities obtained at the Ca\,{\sc ii} triplet remained the ``gold standard" (e.g., \citealt{DroutM33}).  However, with the {\it Gaia} Data Release 2 (DR2, \citealt{DR2}), parallaxes and proper motions for many of the stars in the relevant magnitude range allow us to readily reject most foreground interlopers. 

The second problem has been contamination by these galaxies' own asymptotic giant branch (AGB) stars.  These stars are a late evolutionary stage in low- and intermediate-mass stars, consisting of an inert core, and powered by a helium or hydrogen shell, surrounded by an extended envelope.   \citet{Brunish86} warned that AGB contamination in RSG samples were a potential issue since the higher luminosity AGBs would overlap with lower luminosity RSGs.  In their study of Magellanic Cloud RSGs, \citet{MasseyOlsen} used a luminosity cutoff $\log L/L_\odot$ of 4.9 dex consistent with \citet{Brunish86}'s models, but the exact upper luminosity limit of AGBs are poorly determined.  Some ``super-AGB" stars (the highest mass AGBs) are thought to have luminosities as high as $\log L/L_\odot$ of 5.1 (see, e.g.,   \citealt{2018A&A...609A.114G}.  However, restricting the sample in this way is unappealing, as RSGs of this luminosity are
typically 15$M_\odot$ objects, while the RSG population extends down luminosities of $\log L/L_\odot$ of 4.0 (i.e., $\sim 8M_\odot$).  Fortunately, AGBs and RSGs can be distinguished on the basis of their near-IR {\it J-K} colors, as AGBs are significantly cooler than RSGs of the same K-band brightness, with corresponding redder colors.  This was nicely demonstrated for a sample of red stars in the SMC by
\citet{Yang2019}, following the work of \citet{2006AA...448...77C,2006A&A...452..195C} and \citet{2011AJ....142..103B}, and was used successfully by \citet{LMCBins,UKIRT} in identifying the RSGs of the LMC and two fields of M31.

\section{Observations and Photometry}

Our NIR imaging of M31 and M33 are archival in the sense that in both cases the data were taken for other purposes. We describe the characteristics of each data set below, and summarize some of the basic properties of the surveys in Table~1.

\subsection{M31}
The {\it J} and {\it K$_s$} images of M31 were taken as part of the Andromeda Optical and Infrared Disk Survey ({\sc androids}) aimed primarily at obtaining a uniform surface brightness map across 5 deg$^2$ region of disk and bulge.   The project, calibration, and reduction procedure are described by \citet{2014AJ....147..109S}.

The images were taken with the wide-field NIR camera WIRCam \citep{2004SPIE.5492..978P} on the 3.6-m Canada France Hawaii Telescope located atop Maunakea.  The dataset consists of 70 usable {\it J}- and {\it K$_s$}-band image pairs.  The data were taken in queue mode during four blocks of time:  nine nights during (UT) 2007 Aug 1-Sep 23 (27 fields); 12 nights during 2009 Aug 6-Oct 28 (12 fields); four nights during 2011 Sep 18-Oct 9 (4 fields); and six nights during 2012 Oct 6-Oct 29 (27 fields).  (In addition, there were images obtained
for 3 fields obtained on 2009 Oct 29 and 1 field on 2012-Oct 10 that we judged unusable due to reduction issues.)    Each field was observed in 16 $\times$ 47~s exposures (i.e., 752 s total) in {\it J}  and 26 $\times$ 25~s exposures (i.e., 650 s total) in {\it K$_s$}.  An equivalent amount of time was spent observing sky: because of crowding and the large angular extent of M31, the sky-subtraction was achieved by nodding the telescope away from the M31 field by one or more degrees.  The nodding strategy is discussed
extensively in \citet{2014AJ....147..109S} based upon the first two observing seasons.  A customized 
pipeline, described by \citet{2014AJ....147..109S}, produced the final, stacked images that we used in
our analysis.  They were made available for our project through co-author S. C. 

\begin{figure}
\epsscale{1.3}
\plotone{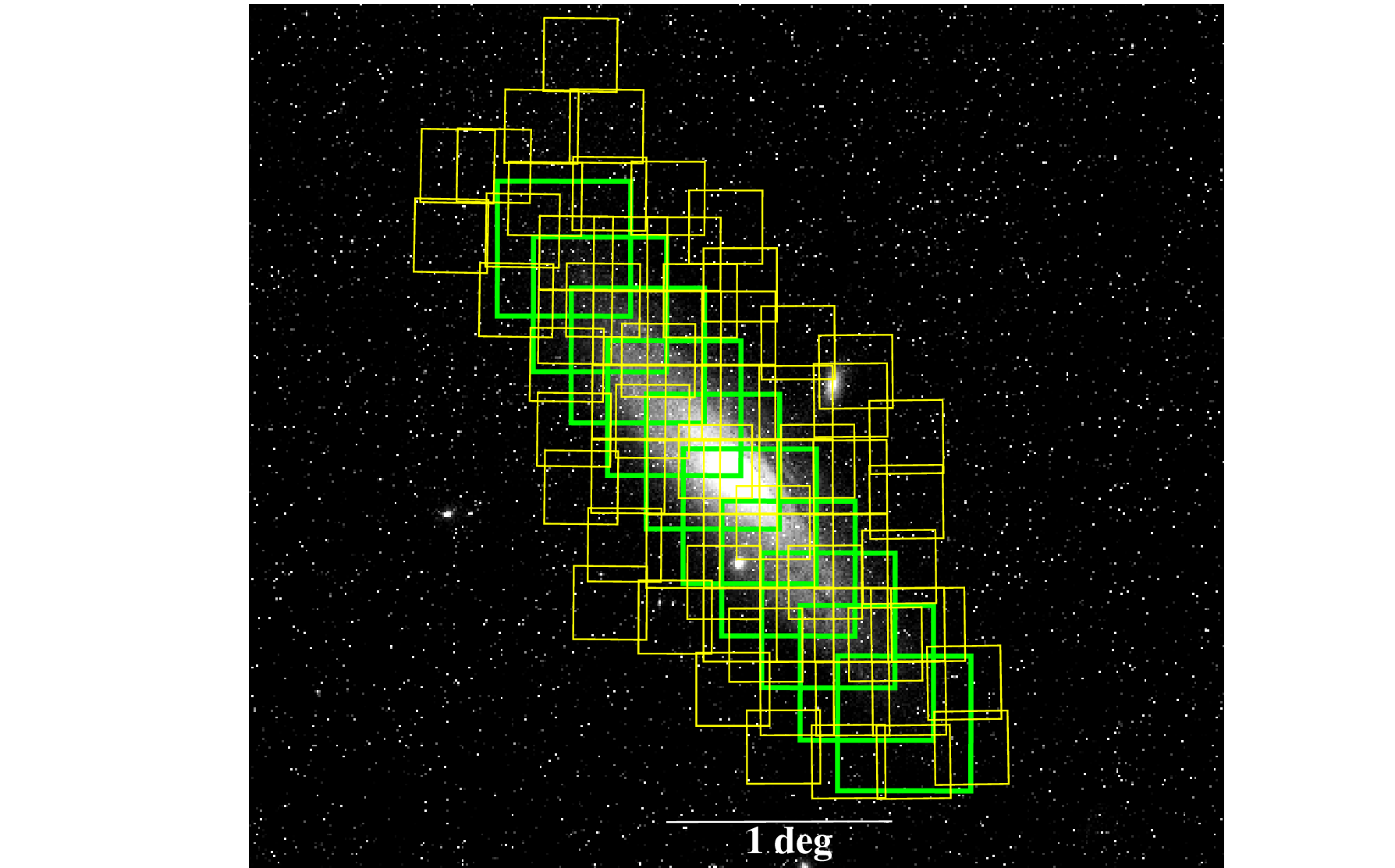}
\caption{\label{fig:m31cfht} M31 survey area.  The outlines of the 70 CFHT fields are shown marked in yellow.  The outlines of the ten fields of the optical Local Group Galaxy Survey are shown for comparison in green.}
\end{figure}

The image scale of each image is 0\farcs3 pixel$^{-1}$.  Although nominally each field covered 21\farcm5 $\times$ 21\farcm5, we found that there were significant issues along the edges, and trimmed each image to the central 4001$\times$4001 pixels, i.e., 20\farcm0$\times$20\farcm0.  This had no effect on the actual areal coverage, given the large overlap between adjacent frames.  Figure~\ref{fig:m31cfht} shows the area surveyed, which includes roughly 5 deg$^2$.  The pipeline had provided an accurate world-coordinate-system, vastly simplifying our task of combining photometry from so many overlapping frames.

Photometry was complicated by the fact that the seeing was so good ($\sim$0.5-0.6\arcsec) compared to the image scale (0\farcs3 pixel$^{-1}$) that the point-spread-functions (PSFs) were generally under-sampled. That, combined with the large quantity of fields, made PSF-fitting an unattractive option.  After numerous experiments, we concluded that we could do little better than standard aperture photometry.  We used {\it daofind} within the  {\sc iraf} environment to identify (mostly) stellar objects 3$\sigma$ above the expected noise based upon the equivalent read-noise and gain.  We then performed aperture photometry with {\it phot} using a 3-pixel (0\farcs9) radius aperture; sky was determined locally using a modal definition in an annulus extending from 3\farcs0 to 6\farcs0. Attempts to centroid before performing the photometry resulted in many problems, primarily moving fainter detections towards brighter neighboring objects, or moving a detection from a stellar source to a nearby diffraction spike from a very bright star.  Therefore we performed the photometry centered on the {\it daofind} positions.

The photometry for each frame and filter were then matched with sources from the 2MASS catalog \citep{2MASS} in order to determine photometric zero-points.   Although the most heavily saturated stars had been masked out in our images by the pipeline reduction, we found that stars that matched with 2MASS sources brighter than $J=14.3$ or $K_s=13.3$ showed symptoms of non-linearity issues. Restricting the sample
to 2MASS stars fainter than these limits, and with photometric ratings of ``A" in the 2MASS catalog, we were then able to calibrate each frame photometrically.  Typically 150-200 stars were used per frame, with zero-point standard deviations of 0.01-0.02~mag, and standard deviations of the mean below 0.005~mag.

The calibrated {\it J} photometry from all 70 fields were then combined and averaged, and the same process repeated for the {\it K$_s$}-band data.   Finally the {\it J-} and {\it K$_s$}-band photometry were matched.  Only stars detected in both filters were retained.  We removed anything fainter than $J=17.5$ or $K_s=19.5$ or
brighter than $J=14.3$ or $K_s=13.3$.  

These bright limits are very similar to what we expect for the brightest RSGs in M31; see, for example, Figure 8 in \citet{UKIRT}.  Nevertheless, we decided to add to this sample any 2MASS stars that were not already in our object list.  In order to allow for the irregular field boundaries, we simply required that a 2MASS star be within 1\arcmin\ of a star with existing photometry.  This added about 6,000 stars, bringing the total to 712,716 stellar sources with photometry.  Of course, we expect only a small fraction (a few percent) to be in the color-magnitude range of RSGs after we remove foreground objects. 

We show the photometric errors as a function of magnitude in Figure~\ref{fig:errors}.   The vast majority of stars have a photometric precision better than 0.02~mags to our faint limit.  Due to the large number of stars (over 700,000) we have plotted only every tenth point for clarity. 

\clearpage
\subsection{M33}

The $J$ and $K_s$ data for M33 were obtained as part of a study of AGB stars in M33 and are described in detail by \citet{2008A&A...487..131C}.  The data were subsequently used by \citet{2015MNRAS.447.3973J} as part of their study of the variability of red giant stars in M33.  

\begin{figure}
\epsscale{1.3}
\plotone{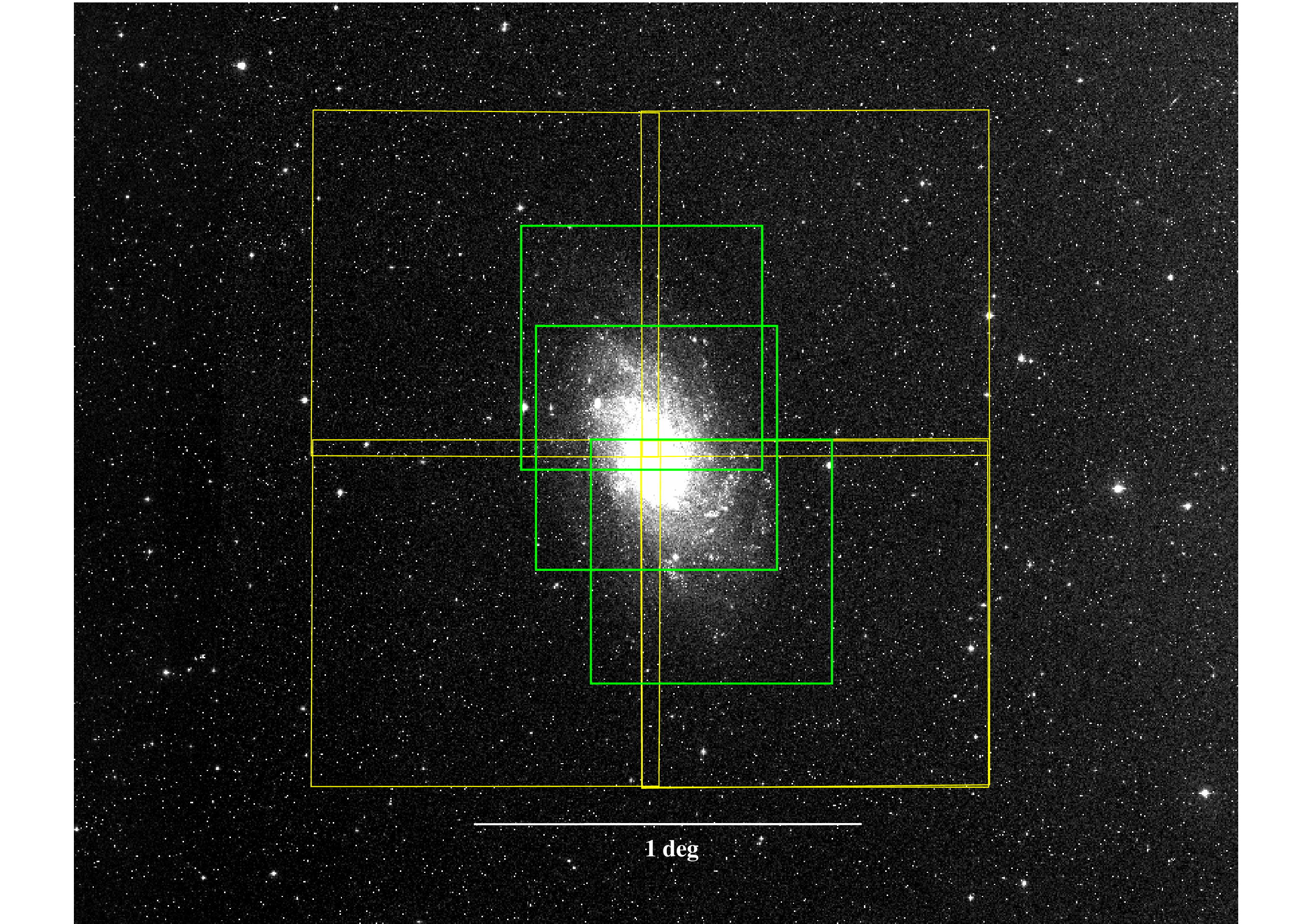}
\caption{\label{fig:m33ukirt} M33 survey area.  The outlines of the 5 UKIRT fields are shown marked in yellow.  (The fifth field is nearly coincident with the other one at lower right.) The outlines of the three fields of the optical Local Group Galaxy Survey are shown for comparison in green.}
\end{figure}

\begin{figure}
\epsscale{0.5}
\plotone{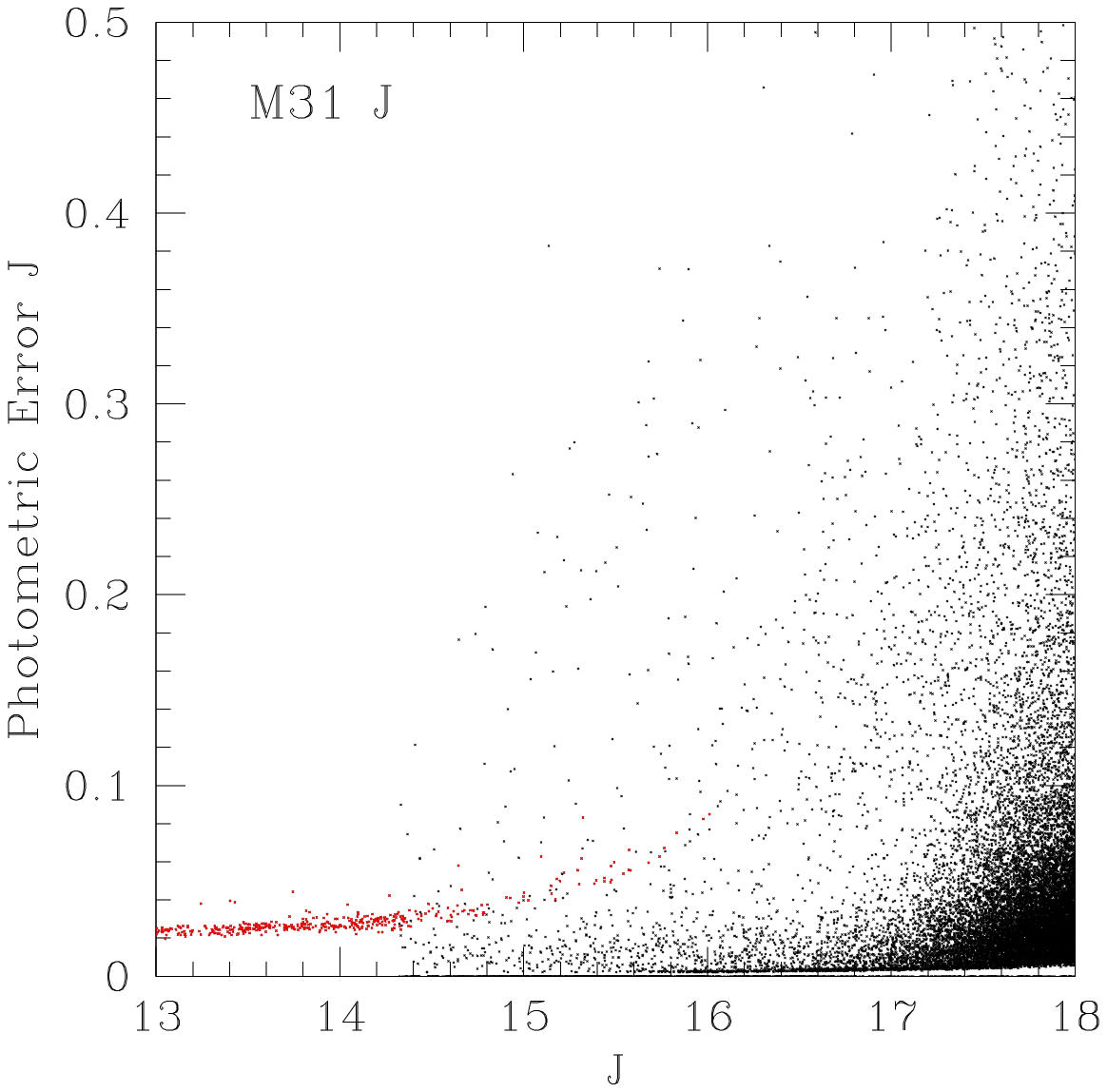}
\plotone{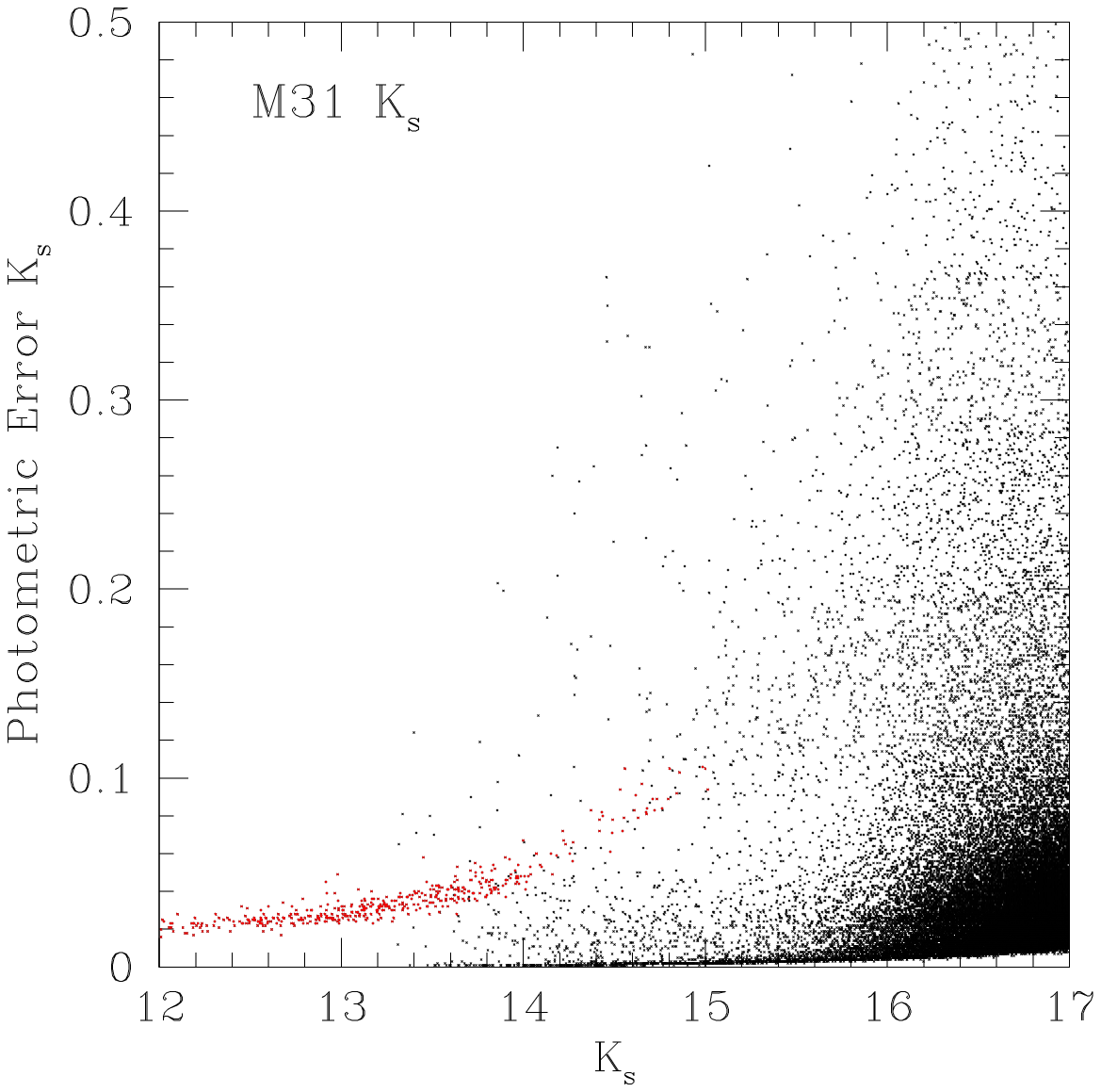}
\plotone{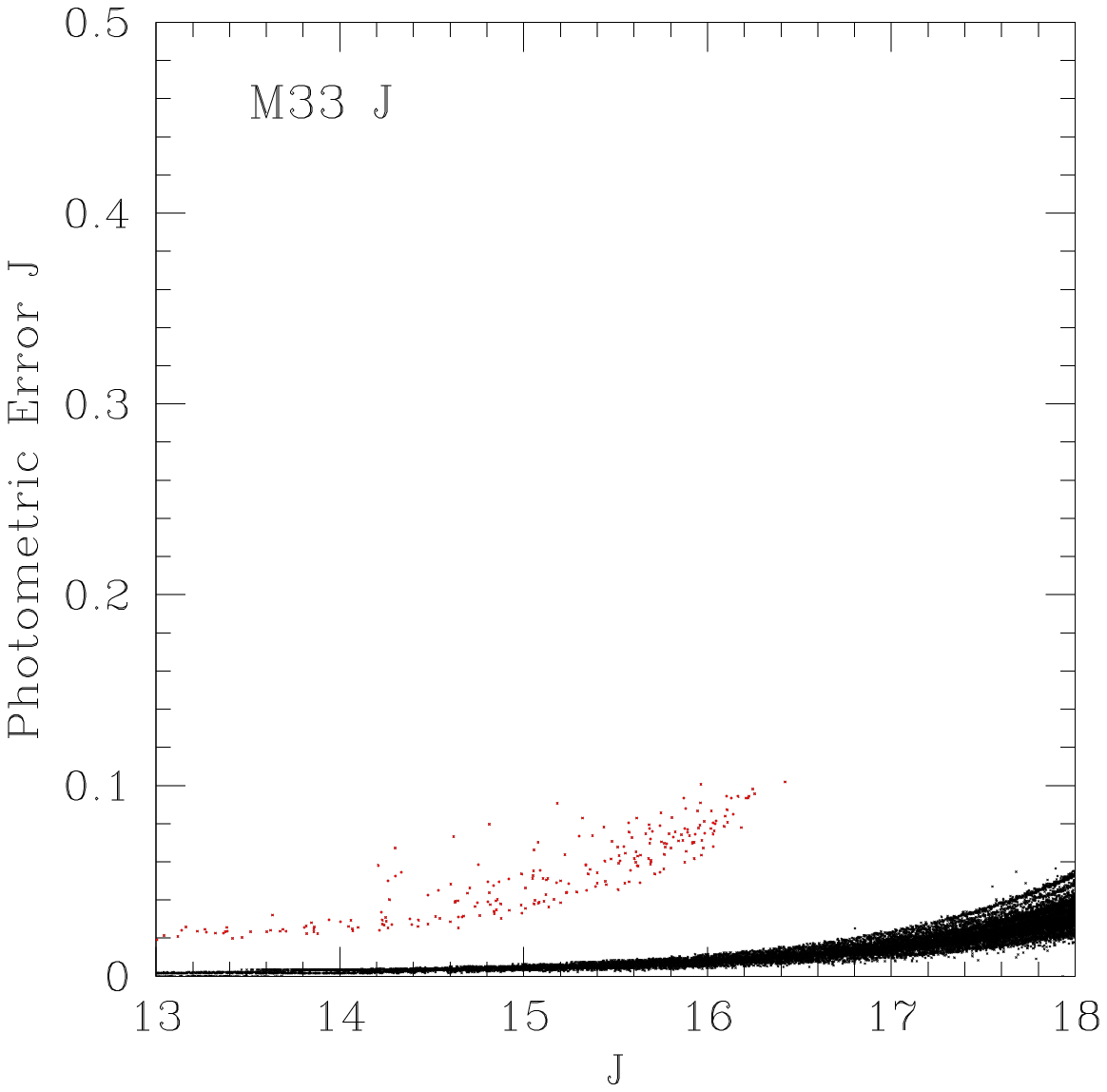}
\plotone{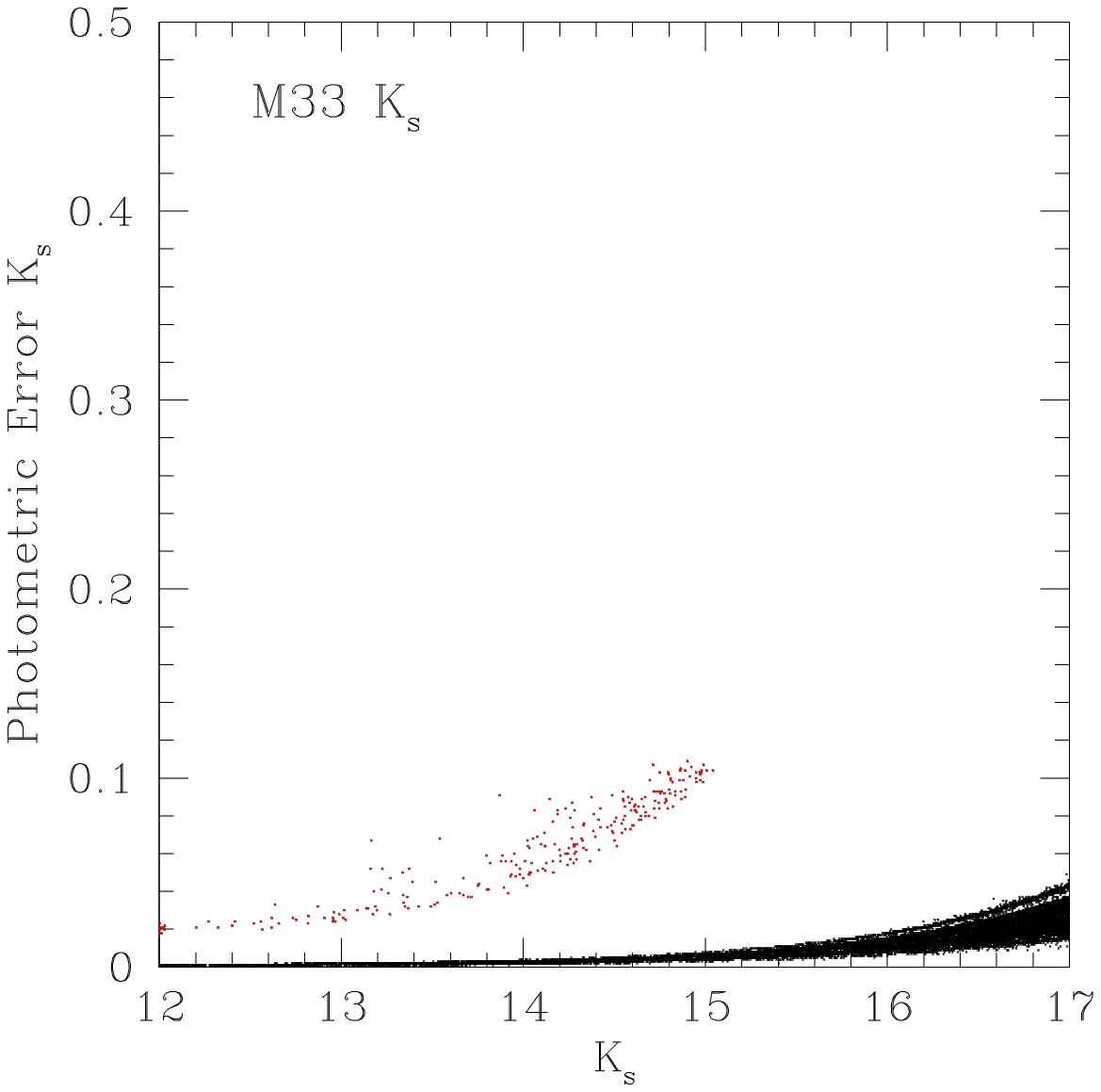}
\caption{\label{fig:errors} Photometric errors.  The photometric errors are shown as a function of magnitude.  The red points indicate the stars that were added from 2MASS.  Note that for clarity we have included only 10\% of the data for the two M31 plots.}
\end{figure}

The data were taken under program U/05B/7 with the Wide Field Camera (WFCam) on the 3.8-m United Kingdom Infrared Telescope (UKIRT) located on Maunakea\footnote{When the data reported here were acquired, UKIRT was operated by the Joint Astronomy Centre on behalf of the Science and Technology Facilities Council of the U.K.}.
The camera consists of four arrays such that with four pointings the gaps between the arrays can be filled in to prove spatial coverage of  52\farcm0 $\times$52\farcm0 with a native image scale of 0\farcs4 pixel$^{-1}$.  (The ensemble of four such pointings is referred to as a ``tile.")  Five such tiles centered on M33 were observed, with some overlap, resulting in coverage of 3 deg$^2$, as shown in Figure~\ref{fig:m33ukirt}.  The equivalent exposure times were 150~s in $J$ and 270~s in $K_s$. (Details of the dithering scheme are given in \citealt{2008A&A...487..131C}.)   The data were taken in queue mode on (UT) 2005 Sep 29-30, Oct 24, Nov 5, and Dec 16.
The data were reduced through the standard Cambridge Astronomy Survey Unit (CASU) pipeline\footnote{http://casu.ast.cam.ac.uk/}, which produces a chip-by-chip source list with $J$ and $K_s$ values \citep{2004SPIE.5493..411I} calibrated from 2MASS along with source classification (i.e., stellar vs.\ non-stellar), and kindly provided to us by the program's P.I., Michael Irwin.  For our analysis we followed the recommendations of the CASU documentation and employed their photometry using a 3\farcs0 radius aperture; these fluxes had been corrected for sky using a locally interpolated value from their
two-dimensional background-following algorithm.

In order to
make sense out of the 160 source lists (4 arrays $\times$ 4 pointing per tile $\times$ 5 tiles in each of the two bandpasses), we first combined all of the measurements on a filter-by-filter basis.  We then averaged the photometry for multiple measurements of the same star.   The averaged $J$ and $K_s$ photometry was
matched with each other, requiring agreement in the coordinates to 0\farcs8 (2 pixels).  In doing so, we paid careful attention to the classification flags.  These are produced by the
CASU pipeline, with values denoting ``stellar," ``non-stellar," ``noise," ``marginal stellar," and ``saturated"\footnote{See https://research.ast.cam.ac.uk/vdfs/docs/catalogues.pdf}.
We found that for objects with multiple detections there was not good consistency between the values
assigned to the same object.  After much experimentation, we found that we obtained the most
satisfactory source list if we required only that an object be classified as ``stellar" or ``marginally stellar"  in
one of the two filters. 

Although the original source lists were nominally calibrated with respect to 2MASS photometry, a comparison of
our final source list with that of 2MASS showed that small corrections were needed to bring them into good
accord; we adjusted the CASU $J$ magnitudes by +0.023~mag and $K_s$ by -0.019~mag.  These values are
comparable to what were found by in analysis of our own UKIRT WFCam data on M31 \citep{UKIRT}.  Similar to that study, stars brighter than $J=12$ or $K_s=12$ began to deviate systematically from the 2MASS values, although they were not flagged as saturated.  Thus, we removed these stars, and added in any 2MASS stars in the same region that were not included in the edited UKIRT photometry catalog.  The final
photometry list included 121,328 stellar objects.  As with M31, we expect only a small fraction to be in the color-magnitude range of RSGs and pass our membership criteria.

We show the photometric errors as a function of magnitude in Figure~\ref{fig:errors}.   M33 is a ``cleaner" field than that of M31, without a large bulge, and with considerably less crowding; hence the photometric errors are smaller, in general.
Again we see that the majority of our photometric errors are $<$0.02~mag to our faint limit.

\section{The Identification of RSGs}

In this section, we proceed to identify RSGs following the same technique as used by \citet{LMCBins} for the LMC and for a previous study by a smaller region of M31 by \citet{UKIRT}.  We begin by using {\it Gaia} to remove many of the foreground stars, and then define a region in the ($J-K_s$, $K_s$) color-magnitude diagram that contains the RSGs but not the AGBs.  Finally, we sanitize the sample by eliminating visually obvious non-RSGs.

\subsection{Identifying Foreground Stars with {\it Gaia}}

Prior to the availabilty of the high-quality proper motions and parallaxes that {\it Gaia} provides, the only way to definitively differentiate between foreground Galactic red stars and M31/M33 RSGs was by the use of radial velocities (see, e.g., \citealt{MasseyRSGs,MasseySilva,DroutM31,Massey2016}). Even so, this method was not effective for stars in the north-east quadrant of M31 as there was too much overlap there between the expected radial velocities of the two groups  due to M31's rotational velocity.  In addition, stars in the Galactic halo will have radial velocities that are similar to the
systemic velocities of these galaxies, as much of that motion is actually the reflex motion of the sun (see, e.g., Figure 5 in \citealt{Stephane}), making it hard to weed out the occasional Galactic halo red giant star.  Broad-band optical photometry also provides a clue, as there is a relatively clean separation between dwarfs and supergiants in $B-V$ at a given $V-R$ \citep{MasseyRSGs} due to the effects of surface-gravity dependent line-blanketing in the blue, although it is clear from radial velocity studies that the method is not completely reliable (see Figure 3 in \citealt{Massey2016}). Fortunately, Data Release 2 (DR2) of {\it Gaia} \citep{DR2} has revolutionized this process. 

Nevertheless, the use of DR2 is not completely straight-forward; to do this properly requires taking spatially-dependent zero-point issues into account as well as considering the intrinsic dispersion in parallax and proper motions within a sample.   As with previous applications, we have followed a process similar to that of \citet{2018A&A...616A..12G}, using a co-variance analysis to separate members from non-members \citep{ErinLBV,UKIRT}, having first established the distributions of parallaxes and proper motions using stars whose {\it Gaia} magnitudes and colors makes it highly likely that they are members.  After cleaning the sample, we find average values for the parallax of -0.033~mas,  and 0.028 mas yr$^{-1}$ and -0.014 mas yr$^{-1}$ for the proper motion in right ascension and declination, respectively, for M31.  For M33 the values are -0.081~mas for the parallax, and 0.103 mas yr$^{-1}$ and 0.034 mas yr$^{-1}$ for the proper motions. With our expected values of proper motions and parallax defined from the median of this sample, we then compare the {\it Gaia} measurements for all the stars in our sample against these values.  A star is considered to be foreground if its proper motion and parallax fall beyond the region that contains 99.5\% of the comparison sample, while a star is considered to be a definite member if it falls within 75\% of the comparison sample.  The membership is considered to be uncertain if the star's value falls between this range. 

However, the \citet{2018A&A...616A..12G} methodology makes sense really only if the membership scatter is dominated by intrinsic dispersion and not the uncertainties in the individual values.   For our M31/M33 sources, that is not the case.  Consider simply the proper motions.  Massive stars in M31 have radial velocities that extend from nearly $-600$ km s$^{-1}$ to $-50$ km s$^{-1}$ due to the galaxy's rotational velocity (see, e.g., Figure 2 in \citealt{Massey2016}). Of course, the space motions perpendicular to the radial velocity will be much smaller than this since M31 is seen mostly edge on (inclination 77$^\circ$ according to \citealt{vandenbergh2000}).  Nevertheless, even if the intrinsic dispersion in the plane of the sky were 500 km s$^{-1}$, at the distance of M31 (760~kpc, \citealt{vandenbergh2000}), this translates to an angular proper motion dispersion of  0.14 mas yr$^{-1}$.  By comparison,
typical uncertainties in the proper motions in our sample are 1-2 mas yr$^{-1}$, an order of magnitude larger.
In other words,  the intrinsic scatter is negligible compared to the individual measurement uncertainties.  This is even more true for the parallaxes, where a 20~kpc long galaxy arranged along the line of sight at the same distance of M31 would have an intrinsic dispersion of 3.5$\times10^{-5}$ mas, compared to a typical uncertainty in the parallax measurements of our stars of 1~mas.

\begin{figure}
\epsscale{0.7}
\plotone{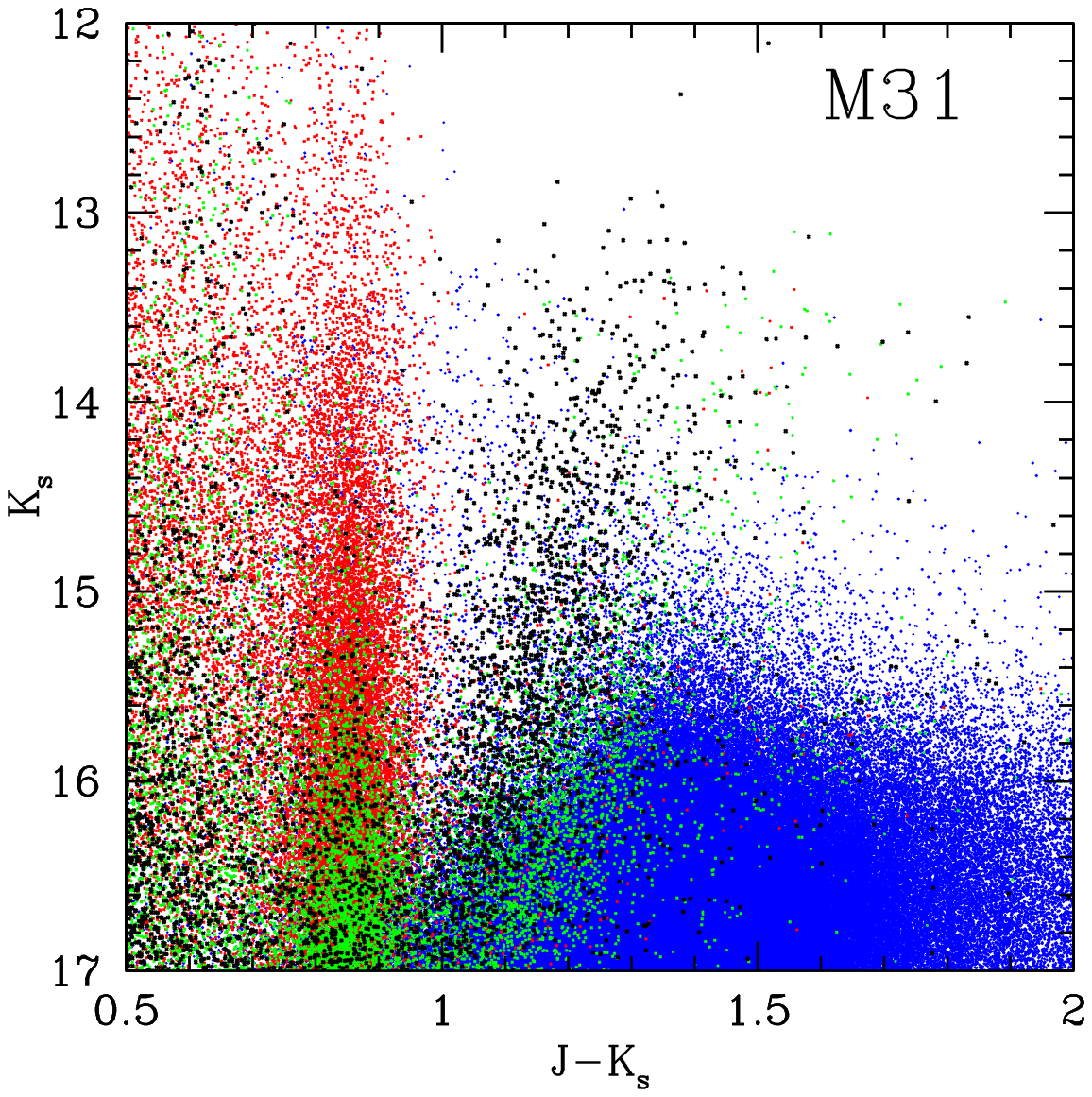}
\plotone{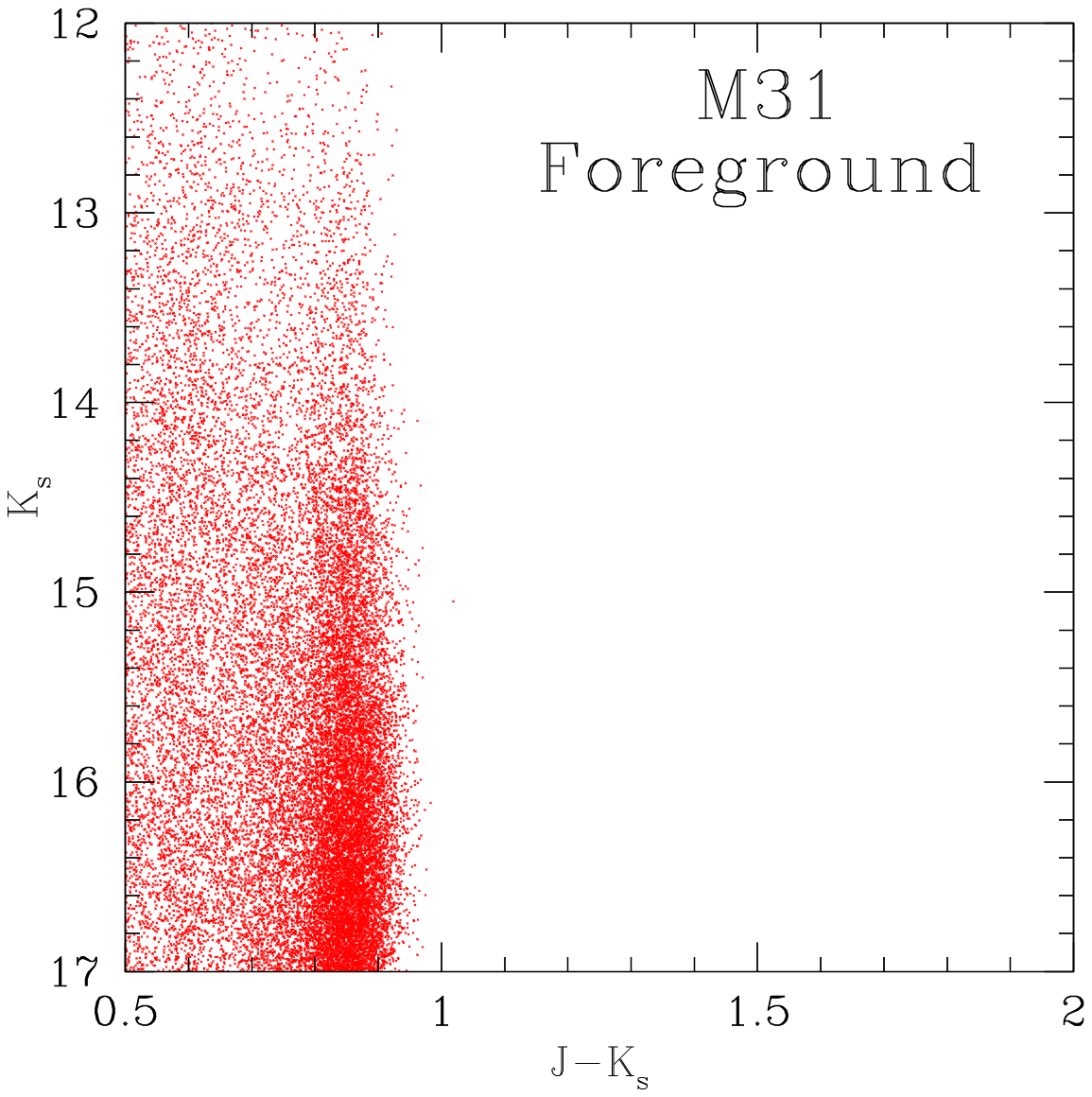}
\caption{\label{fig:M31CMD}  M31 color-magnitude diagram.  {\it Upper:} The color-magnitude diagram is shown for M31, with the
points color-coded depending upon {\it Gaia} determined membership.  The black points are the probable members, while the red points are the foreground stars.  The green points represent the stars with ambiguous membership, while the blue points represent the stars with no {\it Gaia} data. {\it Lower:} The color-magnitude diagram computed using the Besan\c{c}on model of the Milky Way \citep{Besancon} to estimate the foreground contamination.  The simulation covered the same area as our M31 survey, and was centered on M31's location in Galactic longitude and latitude.}
\end{figure}

\begin{figure}
\epsscale{0.75}
\plotone{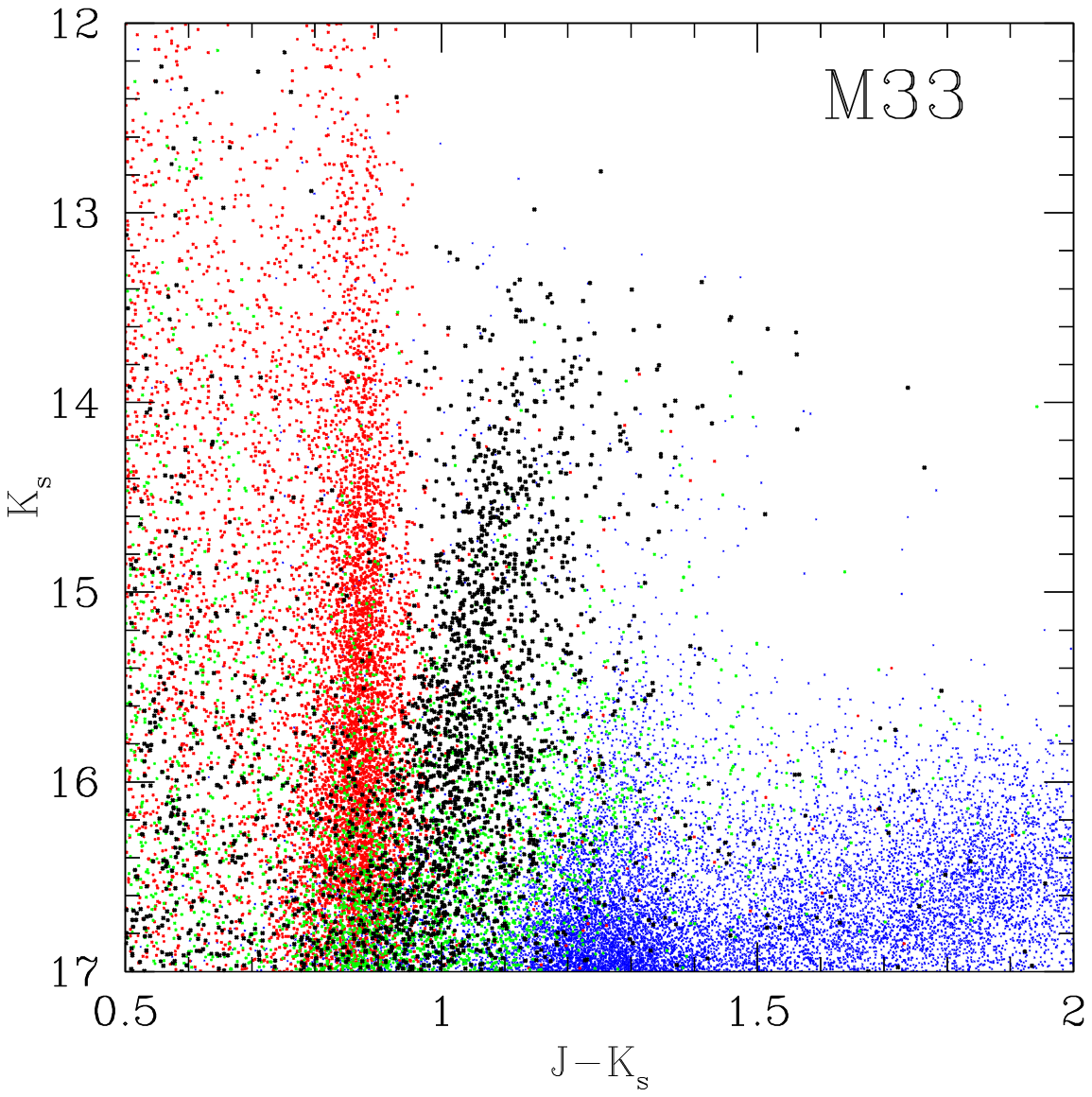}
\plotone{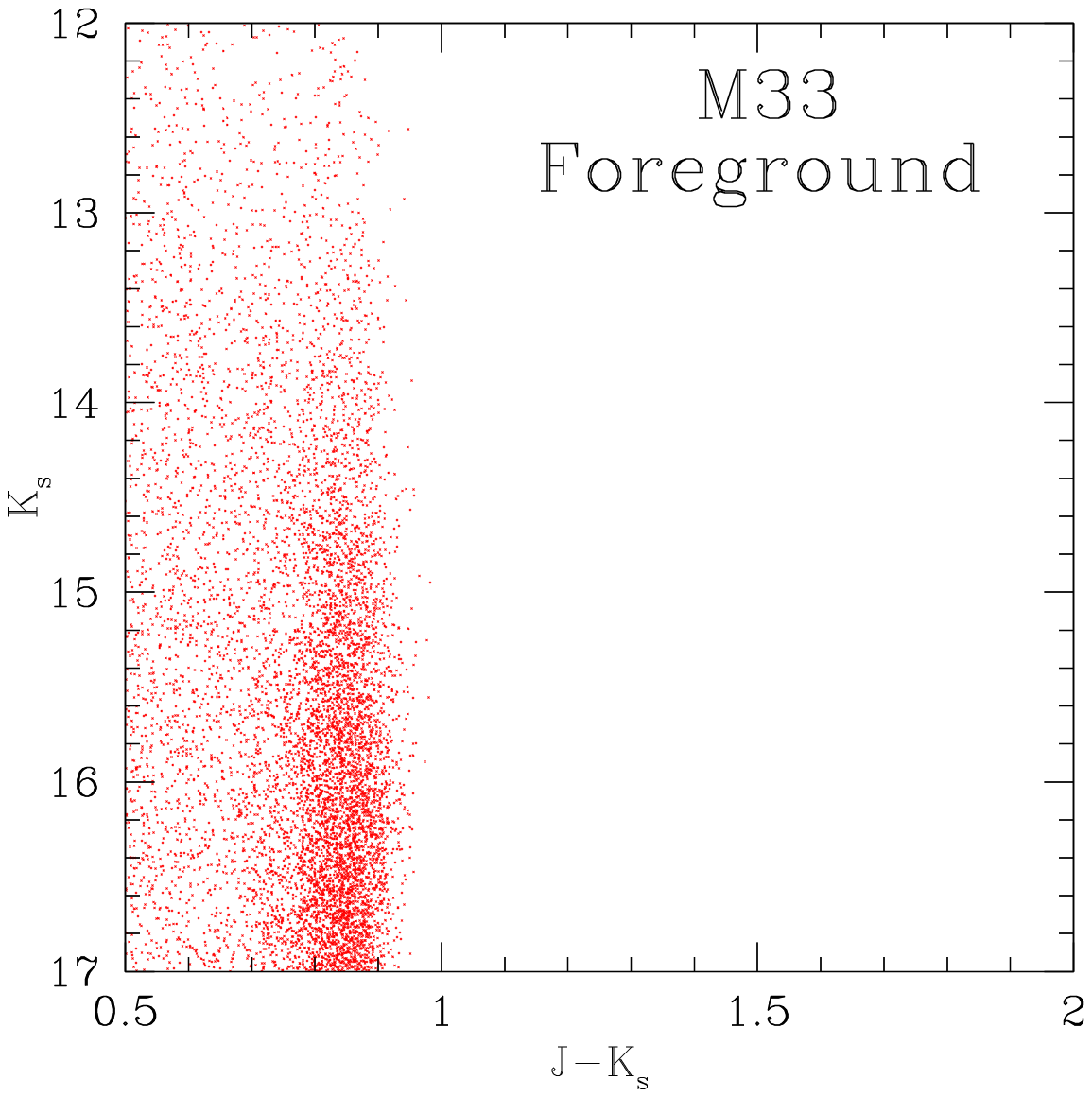}
\caption{\label{fig:M33CMD}  M33 color-magnitude diagram. {\it Upper:} The color-magnitude diagram is shown for M33, with the
points color-coded depending upon {\it Gaia} determined membership.  The black points are the probable members, while the red points are the foreground stars.  The green points represent the stars with ambiguous membership, while the blue points represent the stars with no {\it Gaia} data.  {\it Lower:} The color-magnitude diagram computed using the Besan\c{c}on model of the Milky Way \citep{Besancon} to estimate the foreground contamination.  The simulation covered the same area as our M33 survey, and was centered on M33's location in Galactic longitude and latitude.}
\end{figure}

That being the case, we applied an additional test for stars which the covariance procedure flagged as non-members.  If both the parallax and proper motions were within 2.5$\sigma$ of the expected values (where $\sigma$ is measurement uncertainty), then our classification was changed from ``non-member" to ``uncertain."  We have color-coded the resulting membership determinations in Figures~\ref{fig:M31CMD} and \ref{fig:M33CMD}.   We have also used the
Besan\c{c}on Galactic model \citep{Besancon} to simulate the expected distribution of foreground at the corresponding Galactic longitude and latitudes and covering an area of the same size as our surveys; we see that there is excellent agreement in where in the CMD we find foreground contamination.

The distribution of membership in the CMDs is very similar for the two galaxies.  There is a virtual ``wall" of foreground stars (shown in red) at $J-K_s\sim 0.8-0.9$.  These colors correspond to late K and early M dwarfs  (Table 7.6 of \citealt{Cox}).  At magnitudes fainter than $K_s\sim16-16.5$,  the members (shown in black) and foreground stars begin to blend together, and there are increasing number of stars with ambiguous results (green).  Fortunately this is corresponds to $\log L/L_\odot$ of 4, below which we lose interest.  The triangle of fainter ($K_s<15.5$), redder stars (centered at $J-K_s\sim 1.5$) mostly lack {\it Gaia} data (as indicated by blue), but these are the stars that we expect to be AGBs (see, e.g., \citealt{UKIRT}.)   The comparison with the Besan\c{c}on models confirms that although most of these fainter stars lack {\it Gaia} data, we are justified in our interpretation that these are indeed members.

\subsection{Selecting RSGs from the Color-Magnitude Diagram}

\begin{figure}
\plotone{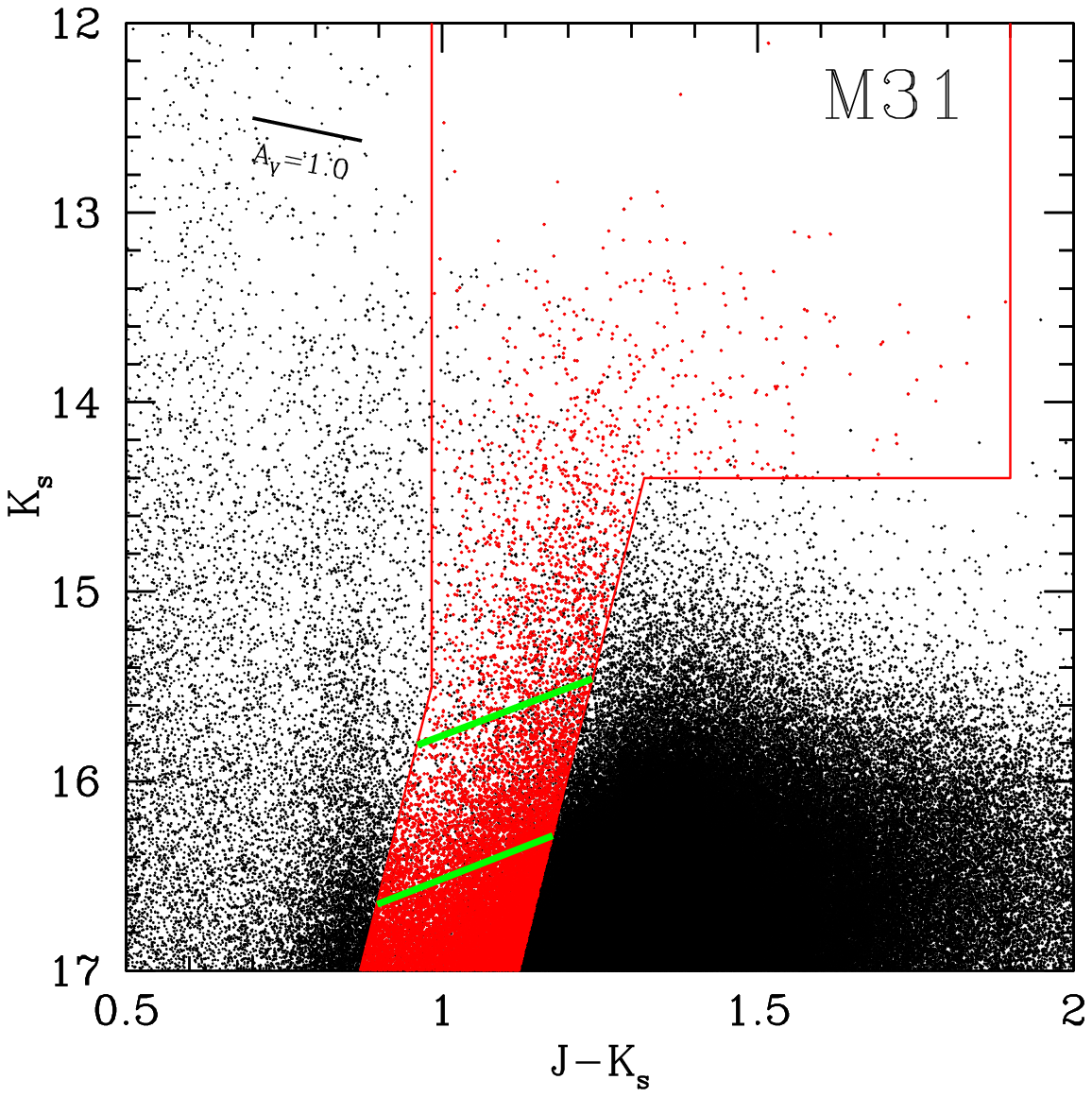}
\caption{\label{fig:M31RSGs} M31 color-magnitude diagram with RSGs identified.  The color-magnitude diagram is shown for M31, with the foreground stars removed and the RSGs region indicated by the red border. The RSG sample of stars is shown by the red points.  The two green diagonal lines show where $\log L/L_\odot=4.0$ and 4.3.  The few points within the RSG boundaries not red are background galaxies or in the regions eliminated as described in the text. The short line at upper left shows the size of a reddening vector corresponding to $A_V=1.0$ mag.}
\end{figure}

\begin{figure}
\plotone{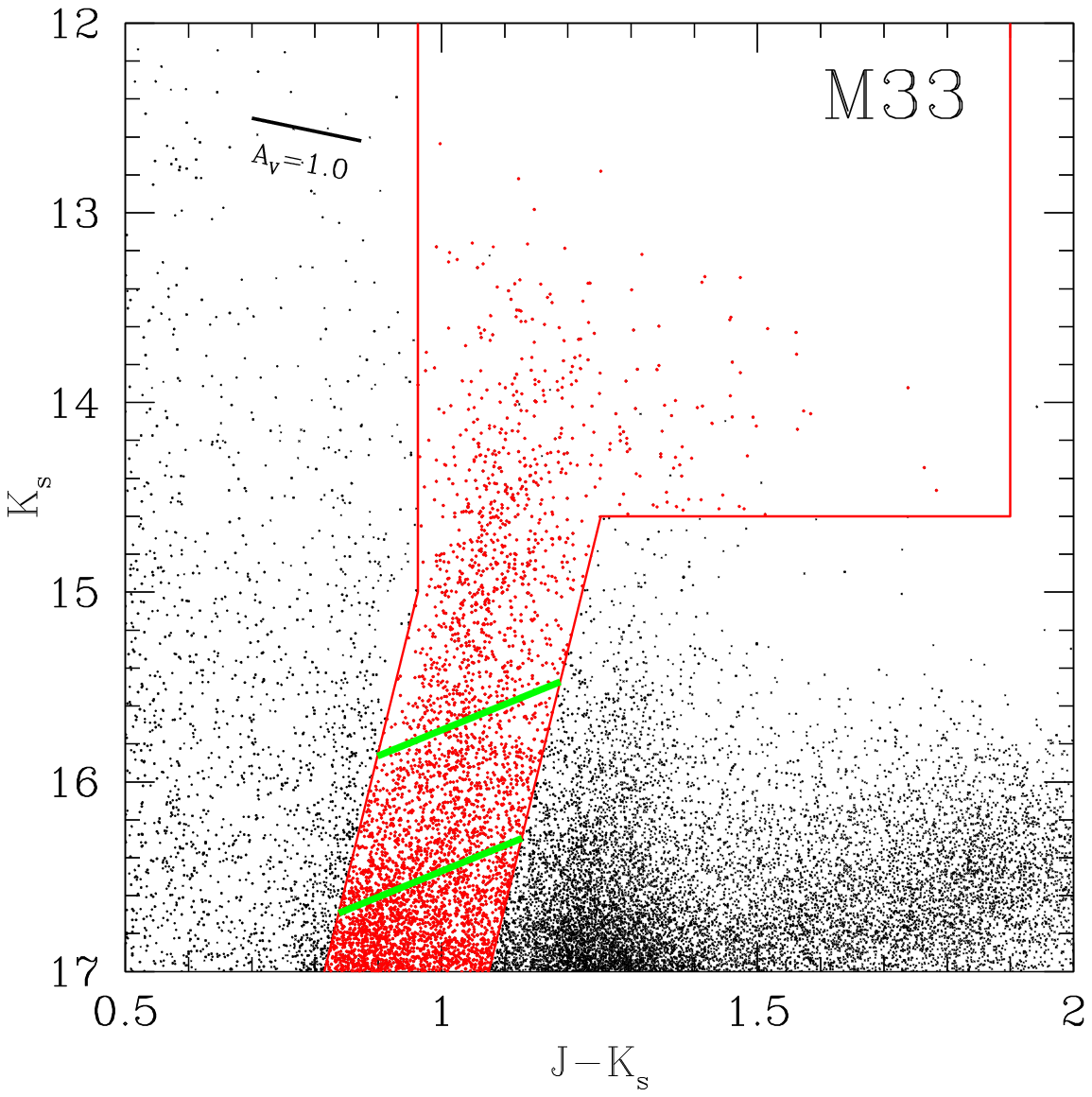}
\caption{\label{fig:M33RSGs} M33 color-magnitude diagram with RSGs identified.  The color-magnitude diagram is shown for M31, with the foreground stars removed and the RSGs region indicated by the red border. The RSG sample of stars is shown by the red points.  The two green diagonal lines show where $\log L/L_\odot=4.0$ and 4.3.  The few points within the RSG boundaries not red are background galaxies.  The short line at upper left shows the size of a reddening vector corresponding to $A_V=1.0$ mag.}
\end{figure}

Following our work in the LMC \citep{LMCBins}, and a small portion of M31 \citep{UKIRT}, we now identify the region in the CMD where RSGs are found, as shown in Figures~\ref{fig:M31RSGs} and \ref{fig:M33RSGs}.
Our ability to separate the RSGs from AGBs stems from the fact that the Hayashi limit shifts to cooler temperatures at lower masses \citep{HayashiHoshi}; thus the AGBs, which are lower in mass than the RSGs, are found at redder colors.  At the same time, the Hayashi limit shifts to cooler temperatures at higher metallicities.  Thus there is more overlap between the RSGs and AGBs in M31 at faint magnitudes than in M33.
We give the adopted boundaries in Table~\ref{tab:FunFacts}, where we have retained the $K_0$ and $K_1$ notation of \citet{2006AA...448...77C}.  

The $K_1$ line denotes the equation separating RSGs from AGBs.  However, for brighter stars, we have expanded the
allowed region to redder colors as there are clearly some stars in this region, and yet these stars are much brighter than
expected for even the brightest AGB stars. (See discussion in \citealt{LMCBins}.)  As described below, we assume these
stars are bright RSGs with extra (presumed) circumstellar reddening.   

In the \citet{Yang2019} study which pioneered this technique to identify the SMC's RSG population, they defined the low $J-K_s$ cutoff line  ($K_0$) as simply parallel the $K_1$ line.  We followed a similar procedure in \citet{UKIRT} in our study of two M31 fields.  However, we have since come to realize that this has no physical basis, and runs the risk of excluding K-type RSGs at brighter magnitudes.  Furthermore, it tends to remove any RSG binaries from the sample, as these will have their colors affected by their bluer companions.
Thus in our analysis of the LMC RSG population \citep{LMCBins}, we used a constant $J-K_s$ cutoff except for the faintest stars, which are still likely contaminated by foreground stars.   We have followed a similar procedure here.

Having make our selection of RSGs based on this CMD, we then performed some manual editing.  Visual inspection of the Local Group Galaxies Survey (LGGS) R-band images \citep{LGGSI} was used to eliminate obvious galaxies, a few of which were misidentified as potential RSGs.  The CFHT data also included M32 and NGC~205, dwarf elliptical companions to M31, visible in Figure~\ref{fig:m31cfht} to the south and north-west, respectively.  We eliminated any spurious detections in the nearby area, i.e, in the rectangular areas
$\alpha$=00:42:28.6 to 00:42:55.0, $\delta$=40:49:25 to 40:54:25 (M32), and  $\alpha$=00:39:55.0 to 00:40:40.0, $\delta$=41:35:10 to 41:48:15 (NGC~205).  There were also numerous detections near the bulge of M31 which were spurious due to the extra noise introduced by the background, and these were also eliminated from our list; we excluded the rectangular region $\alpha$=00:42:30.0 to 00:43:00.5 and $\delta$=41:13:30 to 41:19:00.  M33 had far fewer problems.

The list of RSGs is given in Tables~\ref{tab:M31RSGs} and \ref{tab:M33RSGs} for M31 and M33, respectively. 
We include the IDs, V magnitudes, and any spectral information from the LGGS, using the most recent version \citep{Massey2016}.  There are 38,817 photometrically identified RSGs in the M31 list, and 7,088 in the M33 list. Of these, 31,444 and 4,819 are within their galaxy's Holmberg radius.  As we emphasize below, one must take care to use proper luminosity cuts in discussing the numbers of RSGs.    Only 238 and 197 had been as RSGs in M31 and M33 respectively in the LGGS.   Additional spectroscopy of both galaxies will be discussed in the companion binary frequency paper \citep{M31M33RSGBins}.   Prior to finalizing the RSG lists, we used the matches with the LGGS to remove several QSOs that had been identified in \cite{M31M33QSOs}. In addition, both lists originally contained several WN-type Wolf-Rayet stars; presumably their strong emission lines affected
their $J-K_s$ colors, causing them to appear red.  We removed these, with the exception of J004453.06+412601.7, which is described in the LGGS as ``WN3+M:".  Whether or not this star is a line-of-sight coincidence between a WR star and an M dwarf, or an actual WR+RSG binary, is discussed in the binary paper \citep{M31M33RSGBins}. We also removed two emission-lined luminous blue variables from the lists for each galaxies.  There are a handful of remaining stars whose classifications do not agree with their being RSGs: 32 of the M31 photometrically selected RSGs have previously been been classified as OB stars; there are eight such stars in M33.  In some cases, these are listed as having M-type companions; it is possible that the others are binaries as well.   These need to be re-checked spectroscopically.

 \section{Transformation to Physical Properties and Comparison with Evolutionary Tracks}
\label{Sec-extinct}

Having identified the RSG populations of M31 and M33, we will next use our photometry to determine the
physical properties of effective temperatures and luminosities.  This will allow us to compare the distribution
of stars in the H-R diagram to that predicted by the evolutionary tracks, and at the same time, it will facilitate the future work
we describe in the introduction.  We emphasize that it makes little sense to talk about ``the total number of RSGs" without specifying some luminosity limit associated with that number, and to determine that we must do these transformations.

\clearpage
\subsection{Reddening Corrections}
By far the most uncertain part of this process is what to assume for the extinction towards these stars.  In general, studies of OB stars in both M31 and M33 have revealed only modest reddening.   In many ways, this is most surprising for M31, given that the galaxy is inclined 77$^\circ$ \citep{vandenbergh2000} to our line of sight.  Spectroscopy and photometry of stars in four OB associations of M31 by \citet{MAC86} showed color excesses of $(B-V)$ ranging from 0.08 to 0.24, i.e., $A_V$=0.25-0.75.  Using the LGGS photometry, \citet{LGGSII} used the ``blue plume" of OB stars to estimate the average reddening of OB stars M31, finding a very similar value, $E(B-V)=0.13$ ($A_V=0.4$).  The same method yielded
essentially identical results for M33.   By contrast, an analysis of the Panchromatic Hubble Andromeda Treasury (PHAT) photometry by \citet{PHATDust} suggest an average extinction of $A_V\sim$1~mag for stars in the disk of M31.  We discuss this issue further below, in Section~\ref{Sec-discussion}. 

The reddening situation is further complicated by the fact that RSGs are known to make their {\it own} dust, and in general show circumstellar extinction in excess of that of OB stars in the same region \citep{Smoke}.  Model fitting of a sample of M31's RSGs by \citet{MasseySilva} yielded direct measurements for $A_V$ for a sample of RSGs in M31.  \citet{UKIRT} noted that these results showed a luminosity dependent to the amount of extinction, and adopted $A_V=0.75$ for stars fainted than $K=14.5$, but a linear increase in $A_V$ for stars brighter than that\footnote{Note that although we are conducting our study in the NIR, we discuss the extinction in terms of $A_V$ simply for comparison with other studies; the conversion is given in Table~\ref{tab:FunFacts}.}.   They found that such a procedure was not only consistent with the handful of RSGs with direct $A_V$ measurements, but also resulted in much better agreement with the evolutionary tracks than would have occurred by adopting a constant $A_V$ (see their Figure 10).  Physically this also makes sense: RSGs above some mass limit ($\sim 20M_\odot$) flirt with the Eddington limit during their evolution \citep{Sylvia}, resulting in large episodic mass loss.  

We have used
the same procedure here; the relevant relations are given in Table~\ref{tab:FunFacts}.    In addition, allowing for very
red stars ($J-K_s \sim 1.5$) to be included in the RSG list as long as they were brighter than that expected for AGBs
meant that \citet{LMCBins} had to apply an additional reddening correction for those, with the assumption that their
intrinsic colors were similar to that of other RSGs.  Thus their location the color excess in $J-K_s$ was computed by
moving along a reddening line to center of the RSG distribution.  This resulted in a small number of bright stars having a larger correction for extinction.   We adopt similar relations for M33. The relevant equations are given in Table~\ref{tab:FunFacts}.  We include the adopted $A_V$ values along with our photometry in Tables~\ref{tab:M31RSGs} and \ref{tab:M33RSGs}.

One of the great advantages in conducting our study using NIR photometry is that we are less sensitive to reddening than would otherwise be the case.  The extinction at $K$ is only 12\% that of $V$.  Thus, even a 1~mag uncertainty in $A_V$ translates to an uncertainty of 0.12~mag in $A_K$.  The impact this has on the luminosity determination must take into account the uncertainty in $J-K_s$ and hence in the effective temperature as well.  Using the relations in Table~\ref{tab:FunFacts} we find that $\Delta E(J-K) = 1.46 \Delta A_K$.  For an uncertainty of 0.12~mag in $A_K$ this translates to an uncertainty 0.18 mag in $E(J-K)$.  This would change the
temperature by about 300~K, and the bolometric correction by 0.23~mag.  Thus the overall effect on the luminosity would be 0.15~dex.

\subsection{Transformations to Physical Properties}

There are several steps needed to transform the ($J-K_s$, $K_s$) photometry into effective temperatures and bolometric luminosities.  First, the photometry must be transformed from the 2MASS system to the more
standard \citet{BessellBrett} system, as these will provide the connection with effective temperatures via stellar atmopshere models.  The transformation equations from 2MASS to the standard system are given by \citet{Carpenter}, and repeated in Table~\ref{tab:FunFacts}.  Next, the photometry is corrected for reddening and extinction based upon the adopted $A_V$ values derived as explained above.  The relationships between $A_V$, $A_K$, and $E(J-K)$ come from \citet{schlegel}, and are also summarized in Table~\ref{tab:FunFacts}.  

The basic transformation equations between these intrinsic colors and the desired values for the effective temperatures and bolometric corrections are based upon the MARCS stellar atmospheres \citep{Marcs75,Marcs92,marcs} computed
at various metallicities specifically for the low surface gravities of RSGs by B.\ Plez in our collaborative studies of RSGs in the Milky Way \citep{Levesque2005}, Magellanic Clouds \citep{EmilyMC}, and M31 \citep{MasseySilva}.   A linear equation between intrinsic $(J-K)_0$ colors and effective temperature works very well for RSGs.  There are small differences associated with the metallicity of the models.  Following \citet{UKIRT}, for M31 we adopted a compromise between the solar-metallicity and 2$\times$ solar metallicity models, as a value of 1.5$\times$ solar is more in keeping with recent determinations of the metallicity of young stellar population \citep{Sanders}, with evidence of little or no gradient (see also \citealt{Zaritsky1994}).  M33 has a stronger metallicity gradient, but the absolute values are very poorly established.  Thus we found there was little benefit in providing a correction based on galactrocentric distance. Instead, we adopted the same transformation equation as used in our recent LMC 
study \citep{LMCBins},  corresponding to 0.5$\times$ solar, as that is fairly representative of studies of OB stars \citep{Urbaneja2005}.  (For a few of the differing views of M33's abundances, see \citealt{Kwitter, Zaritsky1994, 2007A&A...470..865M} and \citealt{2011ApJ...730..129B}.) 

By contrast, we find that there is negligible dependence on metallicity in the equation for the bolometric correction  (BC$_K$). Again, the equation is given in Table~\ref{tab:FunFacts}.  Unlike the V-band BCs that many astronomers are familiar with, the K-band BCs have positive values.  Since they are closer to the peak flux of RSGs, K-band corrections also change more slowly than V-band corrections with respect to effective temperature, again minimizing the errors we make by doing our analysis in the NIR. 

Finally we combined the de-reddened K values with the BC$_K$ and the true distance modulus (assumed to be 24.40 for M31 and 24.60 for M33, following \citealt{vandenbergh2000}), yielding the bolometric magnitude.
We then converted these to bolometric luminosities relative to the sun, i.e., $\log L/L_\odot$.

The resulting effective temperatures and luminosities have uncertainties of about 150~K and 0.05~dex, respectively, where we have done error propagation based not only on the photometric errors, but also an uncertainty of 0.5~mag in the adopted values for $A_V$.

The use of a linear relationship with $K_s$ for determining $A_V$ results in a good match with the
evolutionary tracks as we see in the next section.  However, it also has a drawback:  a dozen M31 stars have unlikely large extinction values ($A_V=5-10$) and unrealistically high luminosities.  Several of these stars have ambiguous {\it Gaia} results, but most are considered probable members despite their very bright $K_s$ 2MASS values ($K_s=6-9$).  We have retained these stars in the list. Although they are unlikely to be bona-fide RSGs, they are probably quite interesting IR sources.  None of these overly bright stars are within the field of the LGGS, suggesting they lie well away from regions of star formation.  It is possible that in some cases the membership assessment is wrong.  There is no similar problem for M33.

We note that of the 38,817 photometrically identified RSGs in M31 list and 7,088 in the M33 list,  only 7585 and 3911 (respectively) have luminosities $\log L/L_\odot\geq 4.0$.  Of these, 6412 and 2858 are within their galaxy's Holmberg radius.

\subsection{Comparisons with Evolutionary Tracks}

\begin{figure}
\epsscale{0.7}
\plotone{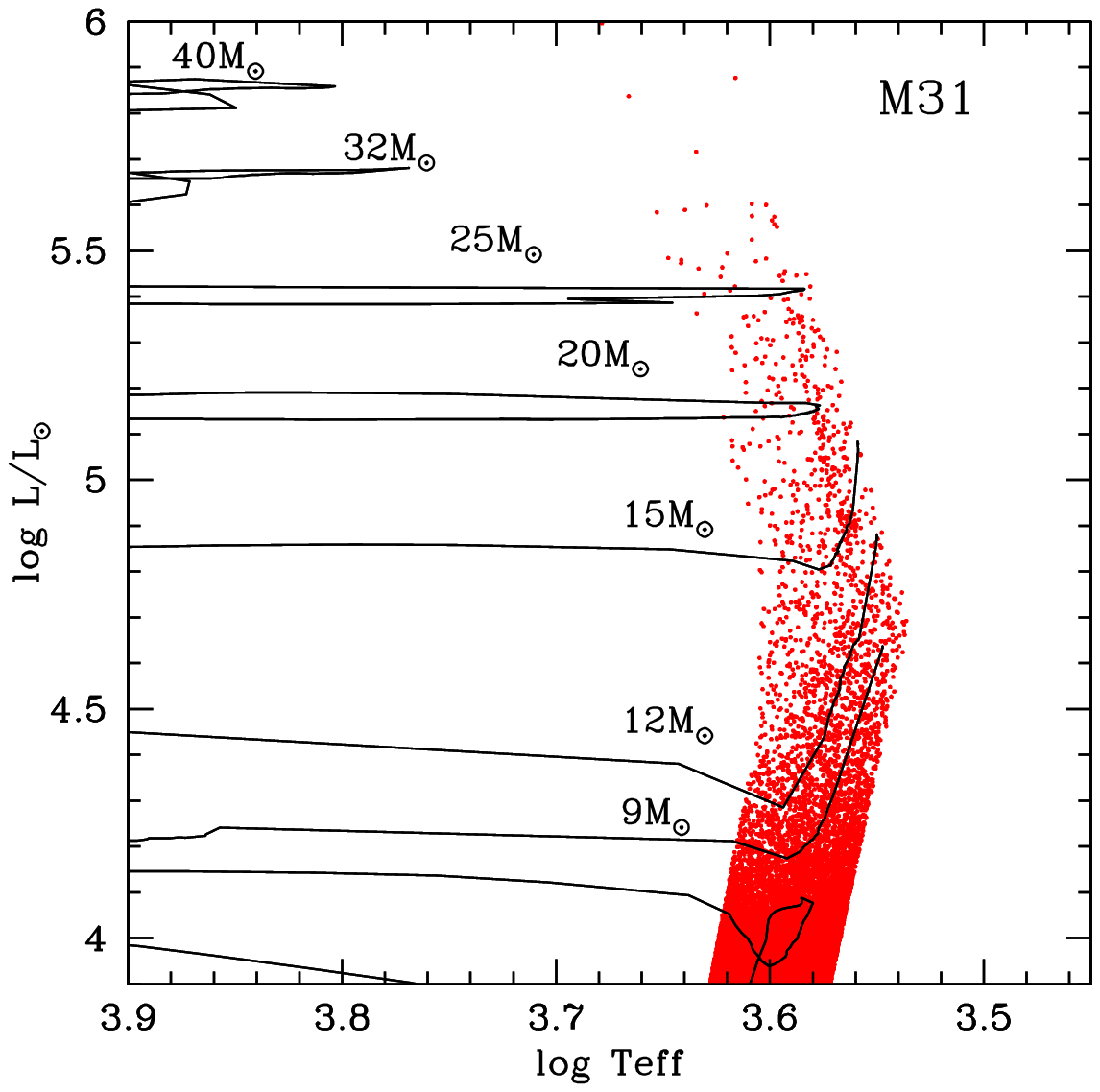}
\plotone{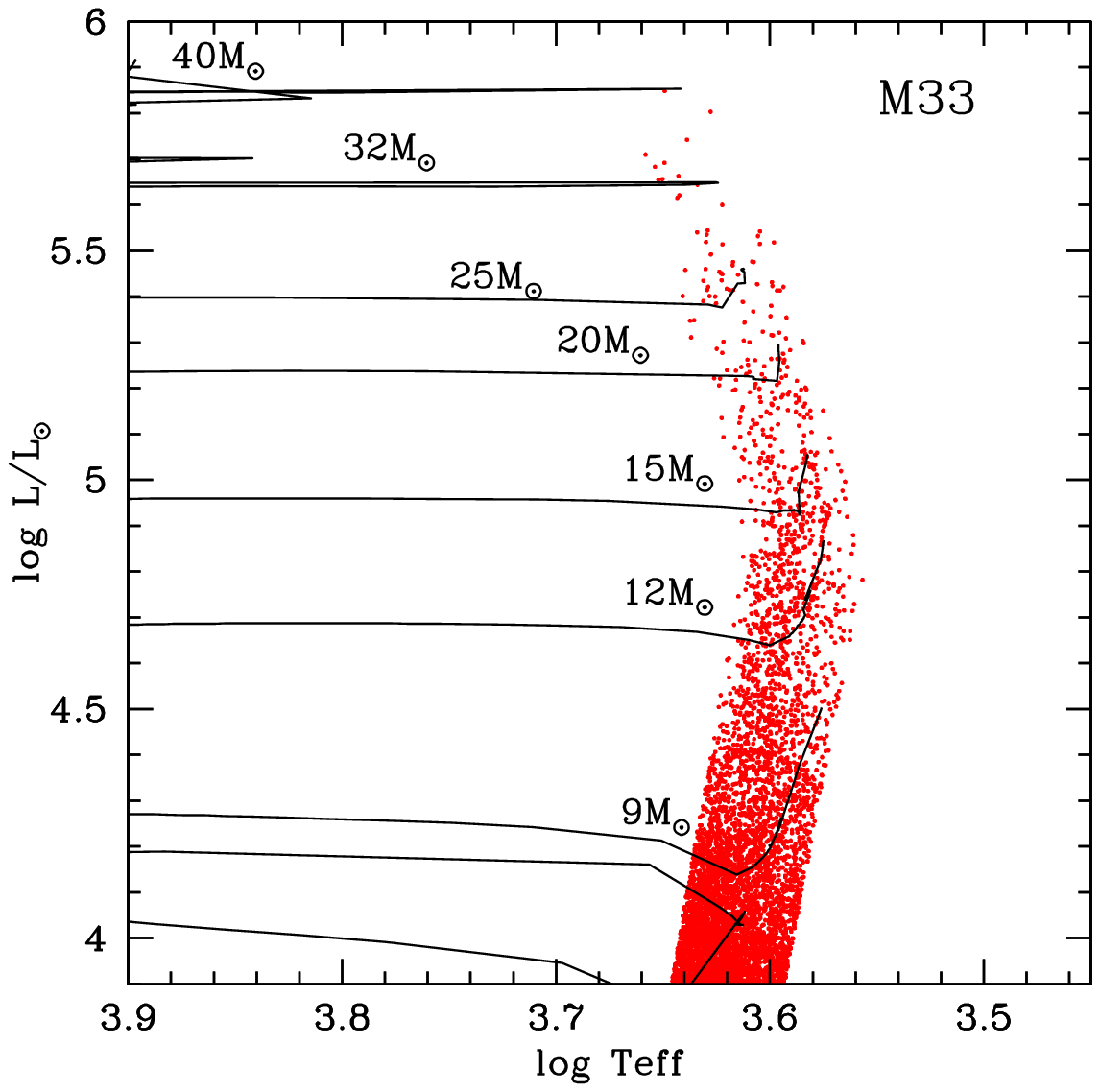}
\caption{\label{fig:HRDs} RSGs in the H-R diagram.  The location of the RSGs relative to those of the Geneva evolutionary tracks are shown.  The M31 plot uses the $z=0.014$ metallicity tracks of \citet{Sylvia}.  The M33 tracks were computed in the same manner by the Geneva group, but correspond to $z=0.006$. There are a few stars with spuriously high luminosity in the M31 sample, as discussed in the text, but generally the distribution of stars matches that expected by evolutionary theory extremely well. }
\end{figure}

In Figure~\ref{fig:HRDs} we show the location of the RSGs in the H-R diagram along with the evolutionary
tracks from the Geneva group.  For M31, we have used the tracks of \citet{Sylvia} computed for solar metallicity ($z=0.014$).
Although a higher metallicity would be more appropriate for M31, such tracks are not yet available.  For M33 we have used the unpublished $z=0.006$ models, kindly made available to us by Cyril Georgy, Sylvia Ekstr\"{o}m, and Georges Meynet. This metallicity corresponds roughly to that of the LMC, and is as appropriate as anything as discussed above.  Although the Geneva models are computed only for single-star evolution, we are using them as the standard for our comparisons as they include the effects of rotation, which is important particularly at sub-solar metallicity, and as their new prescription for the mass-loss rates during the RSG phase have been vetted against observations \citep{UKIRT}.  Binary evolution has little effect on the location of RSGs in the HRD; this is nicely illustrated in the comparison between the single-star Geneva and BPASS binary evolutionary tracks \citep{2016MNRAS.462.3302E,BPASS2} in Figure 8.3 of \citet{LevesqueRSGs}. (See also Figure 7 of  \citealt{2018ApJ...867..155L}.)

The most striking thing about Figure~\ref{fig:HRDs} is how superb the agreement is between the tracks calculated by stellar evolution theory, and the effective temperatures and luminosities derived from our photometry and the MARCS stellar atmospheres.  The upturn and vertical evolution shown by the tracks is due to the increase in the mean particle mass $\mu$ in the stellar cores as He is converted into C and O; the luminosity is a steep function of $\mu$ (see discussion in, e.g., \citealt{LevesqueRSGs}).  Where this happens in terms of effective temperature is dependent upon many things, but particularly the treatment of convection (see, e.g., Figure 9 in \citealt{1987A&A...182..243M})\footnote{We note that all of these complications, and much more, must be dealt with when including binary interactions in evolutionary calculations as well.}.  At the same time, the calculation of the spectral energy distribution in the stellar atmosphere calculations for these bloated stars is fraught with approximations as well.  Thus this agreement between these two complex models seems absolutely remarkable.

The redwards extent of the tracks shift to higher effective temperatures at larger luminosities: at $\log L/L_\odot$=5.5 ($\sim 30M_\odot$) these stars are barely cool enough to be called ``red" supergiants.   This shift is is mostly due to the effects of mass loss during these late stages.  That is why for the solar-metallicity tracks in M31 the 32$M_\odot$ stop at $\log T_{\rm eff}\sim3.77$ (roughly the temperature of the sun!)--mass loss causes the evolution to proceed back to higher
temperatures.  (See discussion in \citealt{Sylvia}.)  At the lower metallicities even
the 40$M_\odot$ models will produce RSGs, as shown by the tracks for M33.   Our data are consistent with these results, although the high-extinction stars mentioned above makes this a little less clean for M31 as we would care.    Additional spectroscopy will resolve this.  We note that
without our luminosity-dependent extinction correction, such agreement would be poorer, as shown in Figure 10 of \citet{UKIRT}.

\section{Resolving the Extinction Puzzle}
\label{Sec-discussion}

We noted back in Section~\ref{Sec-extinct} that both the PHAT survey and our expectations from the M31's inclination would suggest that the average extinction should be higher in M31 than in M33.  However, photometry of OB stars (both in individual associations by \citealt{MAC86} and by use of the ``blue plume" by \citealt{LGGSII}) suggest essentially identical low extinctions.  

As an unexpected bonus, we believe we have solved this puzzle.  In composing Tables~\ref{tab:M31RSGs} and \ref{tab:M33RSGs} we discovered a large discrepancy in the percentages of matches between our NIR-selected RSGs and the LGGs optical catalog between the two galaxies.
For M31 the percentage of RSGs with $\log L/L_\odot>4.0$ that had matches in the LGGS was only 52\%.
By contrast, for M33 the percentage is 87\%. (We restricted the sample the areas which overlap with the LGGS.)  

\begin{figure}
\epsscale{1.0}
\plotone{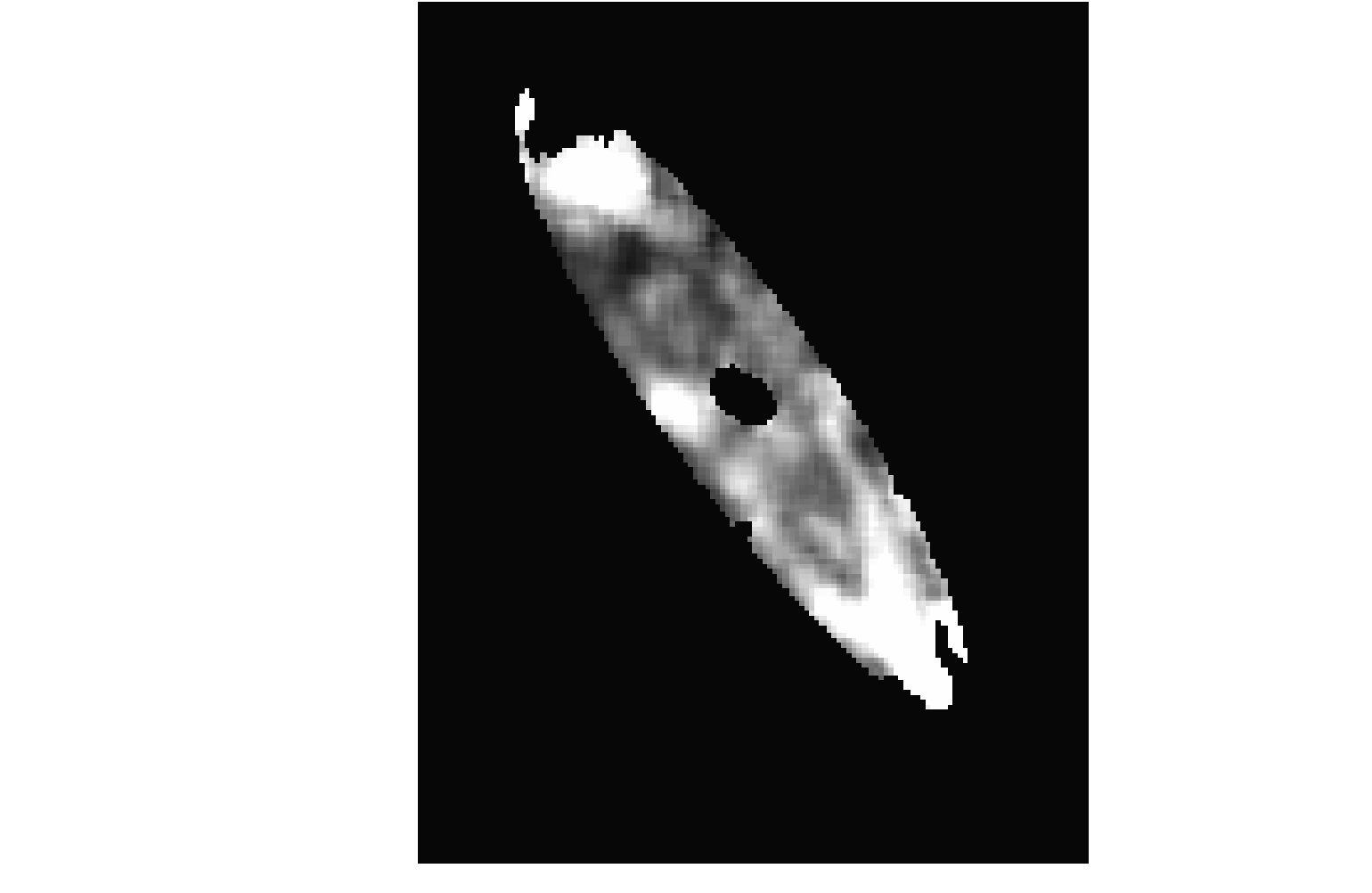}
\caption{\label{fig:matches} Connection between RSGs and LGGS.  This intensity in this image is related to the fraction of matches between our NIR-selected RSG sample the optical counterparts in the LGGS, with brighter regions showing a higher fraction of matches.  The structure is a good match to extinction maps of M31, notably \citet{2014ApJ...780..172D}.  The area covered corresponds to that of the green boxes in Figure~\ref{fig:m31cfht}.}
\end{figure}

We carefully examined this discrepancy and found that the percentage of matches in M31 were similar to that of M33 in the two extreme outlying LGGS fields (the north-eastern and south-western most one) which are sparse,
but after that were mostly poor in all of the other fields.   In Figure~\ref{fig:matches} we show a grey-scale map of the matches.  The field here roughly corresponds to that shown by the green squares in Figure~\ref{fig:m31cfht}.   In constructing this map we smoothed over 5\arcmin $\times$ 5\arcmin\ squares,
and insisted that there be at least 50 stars in each square; if not, the region is black.  The percentage of
matches varied then from 12\% (dark) to 90\% (bright).  We find there is an excellent match with the
dust maps of \citet{2014ApJ...780..172D}; see in particular their Figure 2.  Thus, we  believe this is telling us is that the prominent OB associations and the blue plume stars are the ones generally on the ``near side" of M31's disk, and thus least affected by dust.  This is of course a classic example of Malmquist bias, and explains the apparent discrepancy.

\section{The Spatial Distributions of RSGs}

In an influential study, \citet{HumphreysSandage80} discuss the stellar content of M33, identifying blue and red stars across the face of the galaxy, and defining its OB associations\footnote{Indeed, this paper was instrumental in shaping the course of research for the next 40+ years for P. M., who was an impressionable graduate student at the time. Note that the on-line version does not do the photographic finding charts justice.}.  The blue stars and their associated OB associations clearly followed the spiral arms.  However, the red stars clearly did not.  (Compare their Figure 21 with their Figure 22.)  Although we expect there to be some age difference between the OB stars and RSGs (1-5~Myr vs 10-15~Myr, say), this is not sufficient by far to explain the effect. \citet{HumphreysSandage80} correctly surmised that many of the red stars were foreground dwarfs in our own Galaxy.  (Their photographic study did not go sufficiently deep to detect AGBs.)  This work was followed by large-area photographic surveys of  M33 by \citet{1993ApJS...89...85I}, which suffered from a similar problem.  \citet{MasseyRSGs} devised a method to separate extragalactic RSGs from foreground dwarfs using two-color plots as mentioned above; however his CCD study covered only several small regions of M31 and M33. \citet{2005AJ....129..729R} performed a wide-field CCD study of M33 aimed primarily at  identifying carbon-rich AGBs. No correction for foreground contamination was made.  They identified a region of their CMD that likely contained RSGs, and in their density plot (their Figure 8) indeed a hint of the spiral pattern can be discerned. 

We have the advantage, of course, of being able to use {\it Gaia} to remove foreground stars, although again it should be noted that the majority of the stars in the AGB region of the CMD do not have {\it Gaia} measurements.  However, at these extreme red and faint colors, little foreground contamination is expected, as shown previously in Figures 4 and 5.   We were therefore interested to see what the spatial distribution of RSGs and AGBs looked like in our cleaner samples.

\begin{figure}
\epsscale{0.8}
\plotone{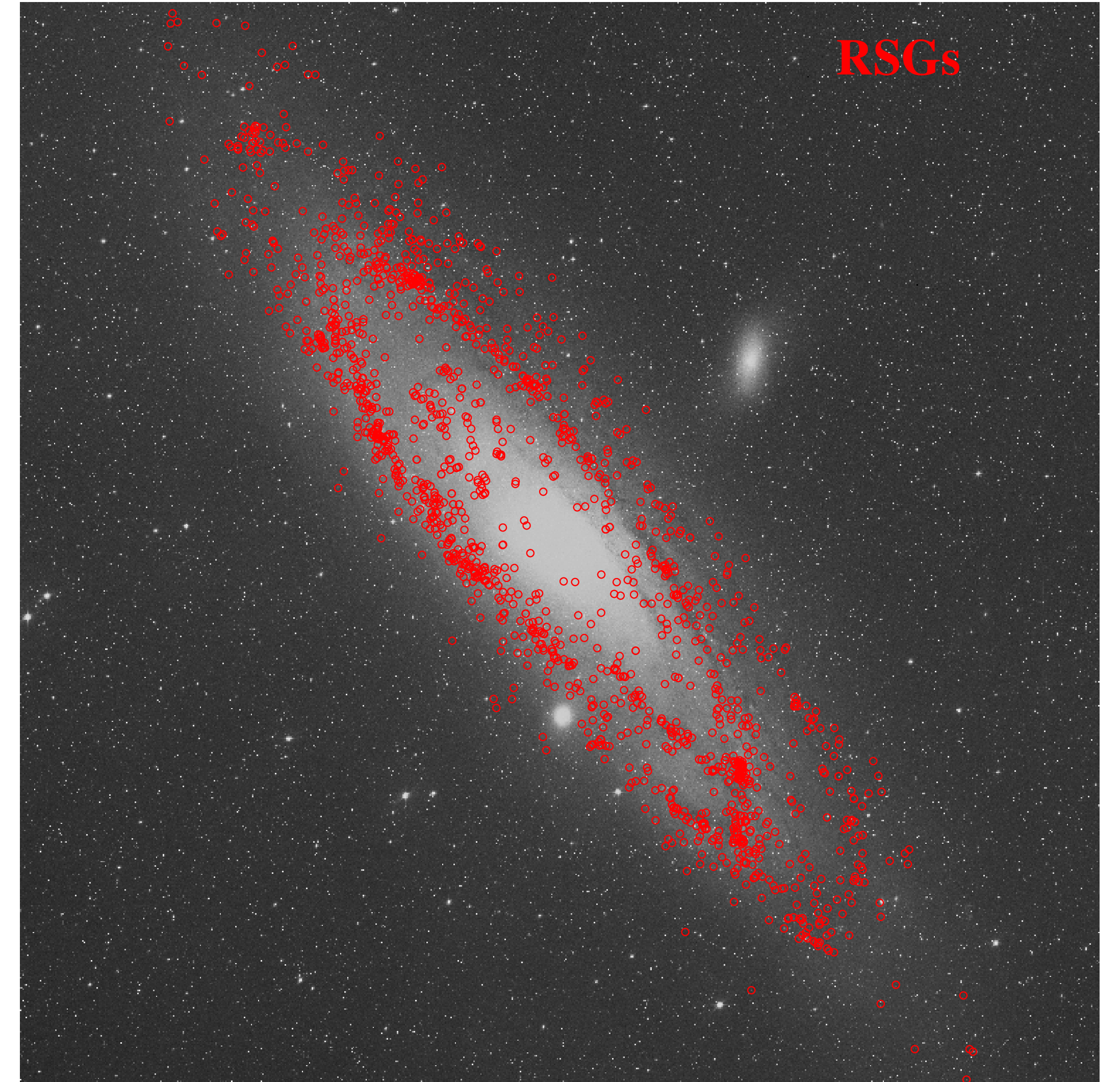}
\plotone{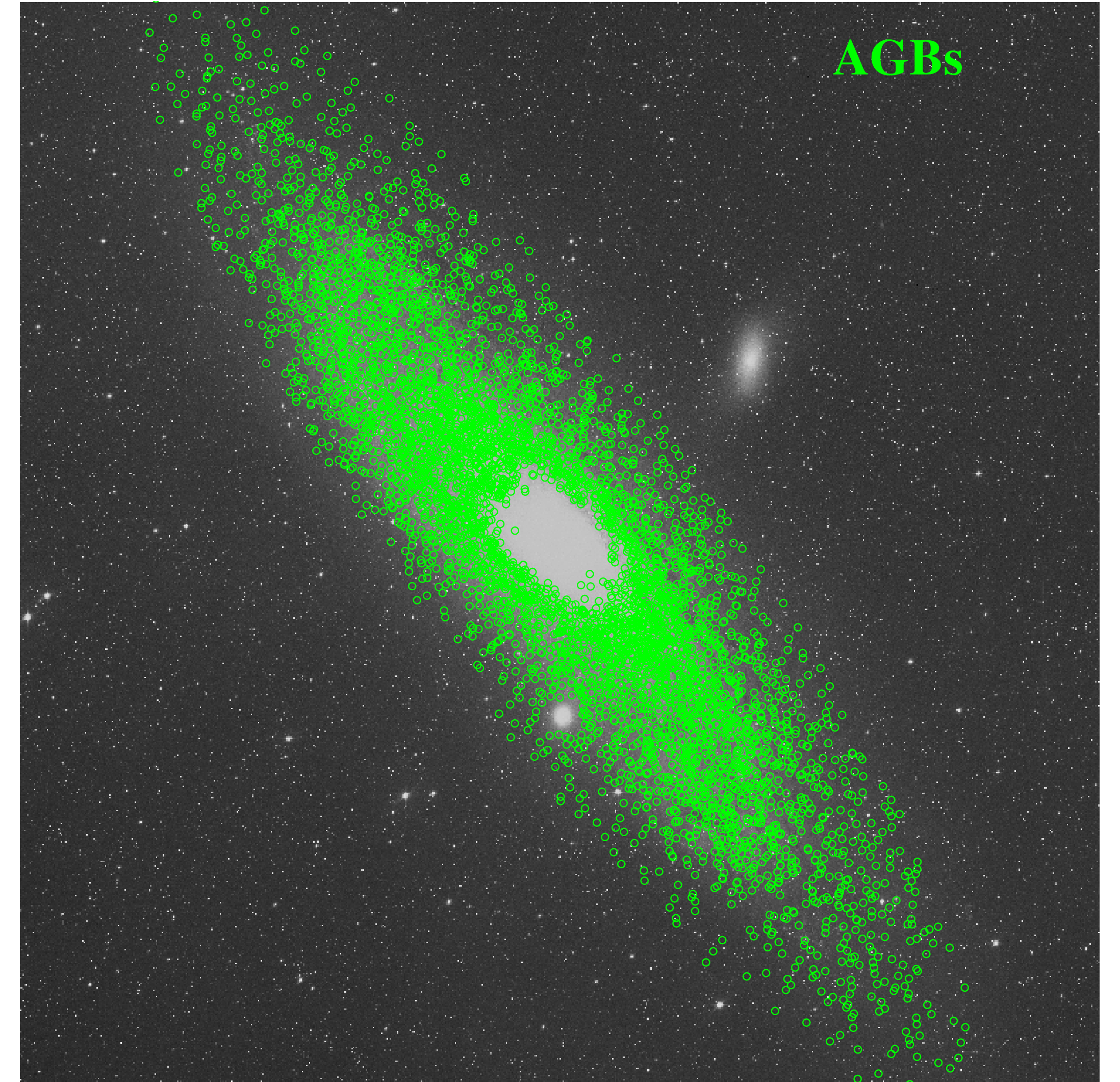}
\caption{\label{fig:M31RSGAGB} Distributions of RSGs and AGBs in M31.  We compare the spatial distributions of stars identified as RSGs and AGBs within the Holmberg radius of M31.  {\it Upper:} We have restricted the sample to those RSGs with $\log L/L_\odot>4.3$ (roughly corresponding to $K_s<15.5$; see Figure~\ref{fig:M31RSGs}).  {\it Lower:} The AGBs are the stars redder than the RSGs sample in Figure~\ref{fig:M31RSGs} and with $K_s<17.0$.  For clarity only 2\% of the stars in our photometric sample are plotted in the lower diagram.  Note that we have excluded the region containing M32 and the nuclear region of M31 as described in the text.} 
\end{figure}

\begin{figure}
\epsscale{0.8}
\plotone{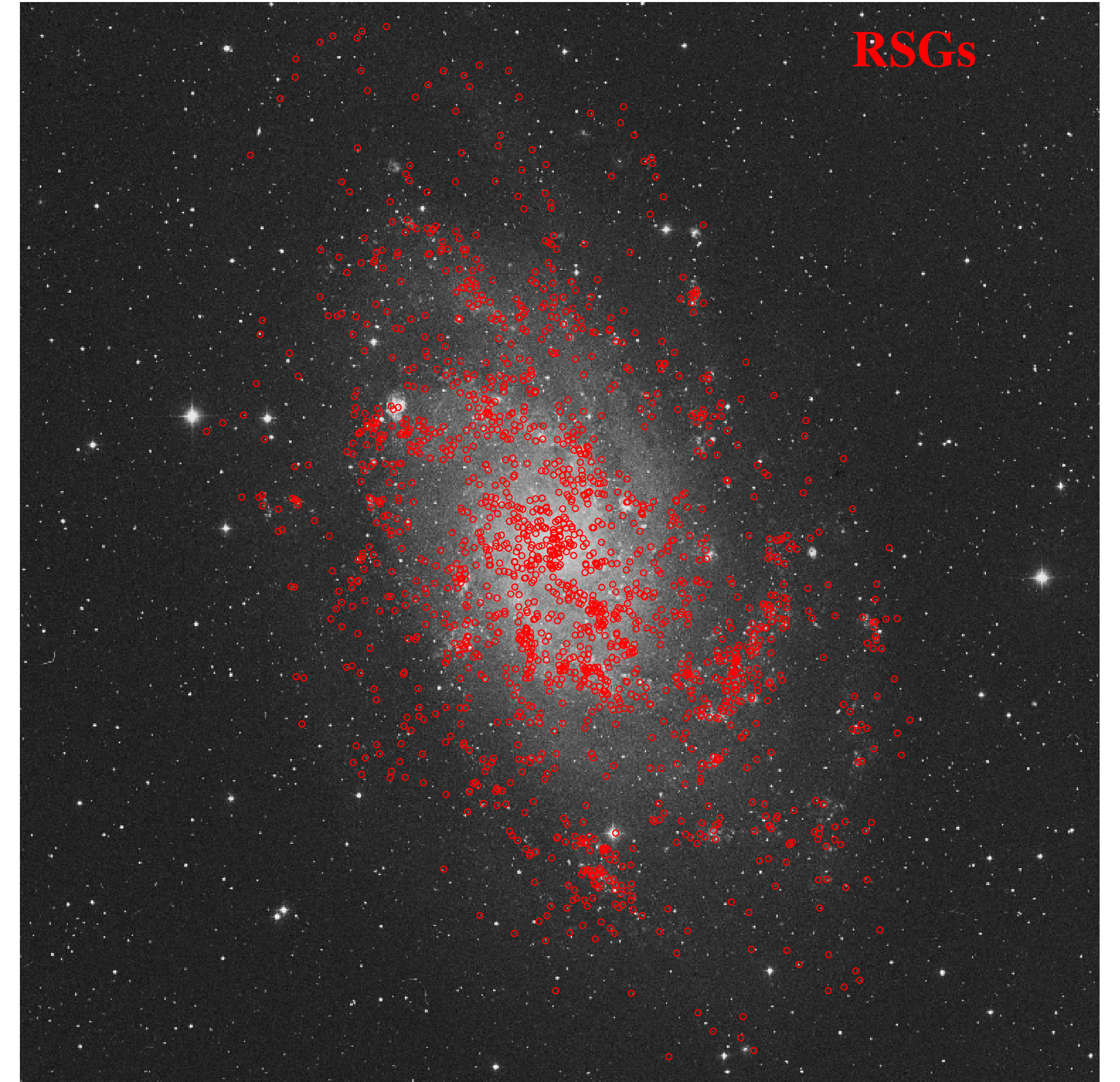}
\plotone{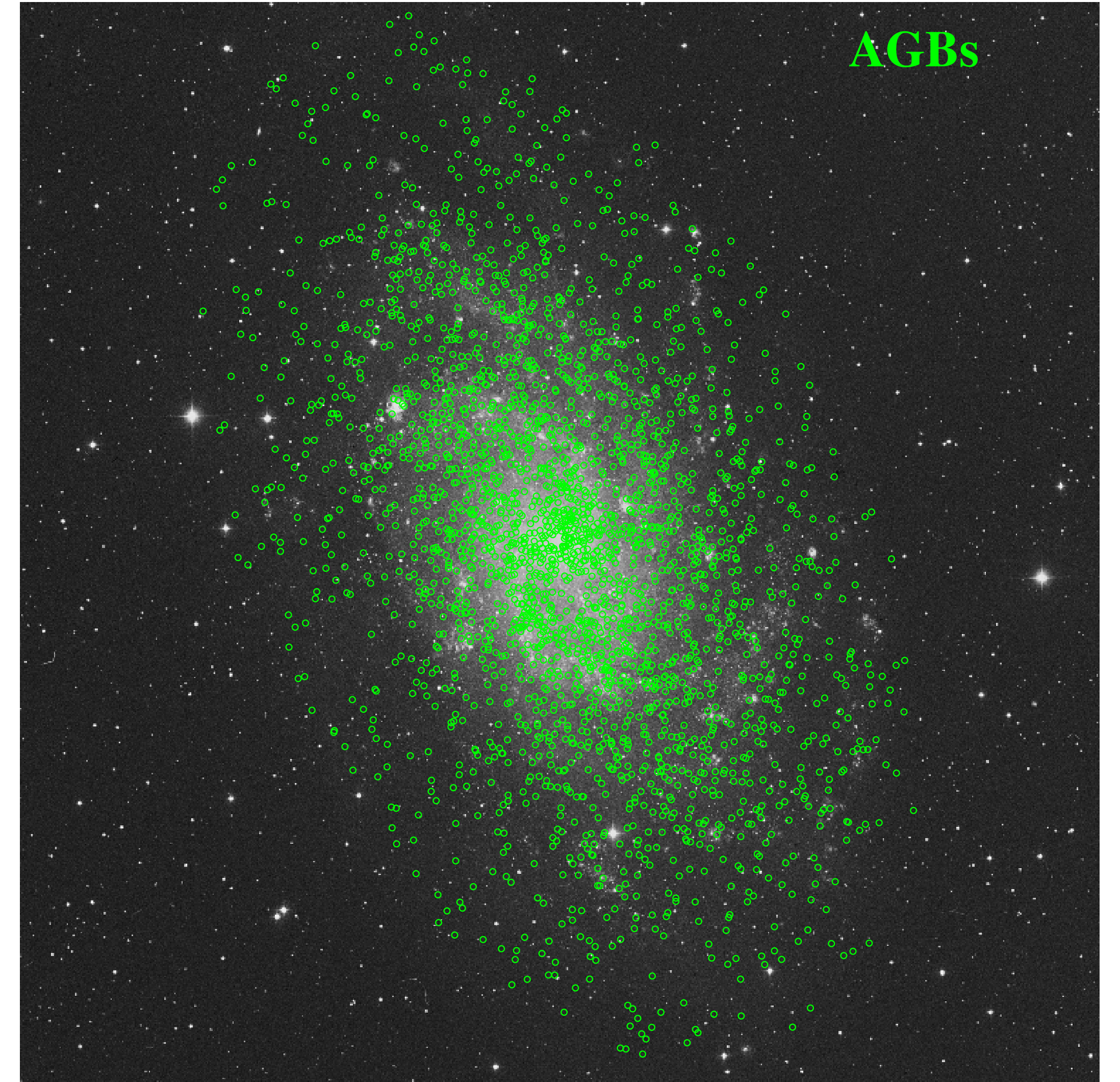}
\caption{\label{fig:M33RSGAGB} Distributions of RSGs and AGBs in M33.  We compare the spatial distributions of stars identified as RSGs and AGBs within the Holmberg radius of M33.   {\it Upper:} We have restricted the sample to those RSGs with $\log L/L_\odot>4.3$ (roughly corresponding to $K_s<15.5$; see Figure~\ref{fig:M33RSGs}).  {\it Lower:} The AGBs are the stars redder than the RSGs sample in Figure~\ref{fig:M33RSGs} and with $K_s<17.0$.  For clarity only 20\% of the stars in our photometric sample are plotted in the lower diagram.}
\end{figure}

The results are shown in Figures~\ref{fig:M31RSGAGB} and \ref{fig:M33RSGAGB}.  In making these figures, we have restricted the sample to stars within the Holmberg radius ($\rho<1.0$).  For the RSGs we have imposed $\log L/L_\odot>4.3$ (roughly corresponding to $K_s<15.5$; see Figures~\ref{fig:M31RSGs} and \ref{fig:M33RSGs}), and for the AGB stars $K_s<17.0$.  (Even so, only a small subset of the AGB stars are plotted for clarity; 2\% for M31 and 20\% for M33.)  As expected, the RSGs very much follow the spiral arms.  In M31 (Figure~\ref{fig:M31RSGAGB}) the vast majority of RSGs are found in the well-known star-formation ring where most of the HI,  OB associations, and H$\alpha$ emission are located (see, e.g.,  \citealt{1966ApJ...144..639R,vdbergh,1994AJ....108.1667D}, respectively).   This suggests that despite our finding in the previous section,
huge sections of M31's massive star population do not remain hidden. In M33 (Figure~\ref{fig:M33RSGAGB}) the RSGs very nicely follow the spiral arm pattern.  By contrast, the AGBs are more uniformly distributed across the disk.  This is as expected given the extreme difference in ages between these two populations, with the RSGs representing 10-20~Myr old, and the AGBs (100~Myr-1~Gyr for the dominate population; see \citealt{2014A&A...565A...9S}).

\section{Summary and Conclusions}

Our study has identified likely RSGs throughout M31 and M33, using archival NIR data that extend well beyond each galaxy's Holmberg radius.   Using a cut of $\log L/L_\odot$ of 4.0 and restricting the area to within the Holmberg radius of the disk ($\rho\leq1$ in Tables~\ref{tab:M31RSGs} and \ref{tab:M33RSGs}), we find  
6412 RSGs in M31 and 2858 RSGs in M33.   By contrast, only a few hundred RSGs are known within our own Galaxy \citep{LevesqueRSGs}.  Many of these were found through the the HD catalog plus other objective prism studies (e.g., \citealt{1954ApJ...120..478N,1957ApJ...125..408B,1992AJ....104..821M}; see also \citealt{1978ApJS...38..309H}). More recently,
clusters rich in RSGs have been found near the Galactic center; i.e, \citet{2006ApJ...643.1166F,2007ApJ...671..781D,2010A&A...513A..74N,2011A&A...528A..59N,2016csss.confE.120M}.   Further studies are being carried out in other directions (e.g., \citealt{2018MNRAS.475.2003D}), but like all Galactic studies, extinction prevents a true Galaxy-wide census, and uncertainties about distances (needed to distinguish RSGs from giants) persist even in the {\it Gaia} era.   Various authors have suggested that the
Milky Way's RSG population should number in the thousands (e.g., \citealt{1987PASP...99..453G,1989IAUS..135..445G,2012A&A...537A..10M}).
The number of RSGs we've discovered in M31 and M33 gives credence to these predictions.

Using the MARCs stellar atmosphere models, we have computed effective temperatures and bolometric luminosities for all of the stars in our sample, and compared those to the location of the Geneva evolutionary tracks.  The agreement is spectacular, and gives us high confidence in the tremendous progress that has been made in both fields.

A comparison of the M31 RSG sample with the LGGS resolves the discrepancy between the average extinction in M31's disk
estimated from thermal dust emission with the comparatively low reddening found in several OB associations and from the blue plume.  The later samples are likely due to stars on the near side of M31's disk, and suffer lower extinction on average.  Nevertheless, RSGs generally are primarily found in the region where the OB associations and other star-formation indicators are prominent in M31, suggesting that this is not too serious a problem in terms of identifying most of M31's massive star population.  In M33 the RSGs also delineate the spiral arms.  As expected, the fainter, redder AGB stars are much more evenly distributed across the face of these galaxies.

The identification of these RSG candidates will hopefully provide a valuable resource for spectroscopic studies for years to come.   Empirical confirmation of our assumed extinction as a function of luminosity will require well-fluxed spectrophotometry similar to what we have employed in the past.   Of more immediate use, our catalog of RSGs will be the basis for the contemporaneous study of the RSG binary frequency \citep{M31M33RSGBins}, and the a detailed comparison of the relative number of WRs and RSGs with stellar evolutionary models in an upcoming paper.

\acknowledgments
Lowell Observatory sits at the base of mountains sacred to tribes throughout the region. We honor their past, present, and future generations, who have lived here for millennia and will forever call this place home.
 Similarly, K. F. N. and E. M. L. would like to acknowledge that they work  on the traditional land of the first people of Seattle, the Duwamish People past and present, and honor with gratitude the land itself and the Duwamish Tribe. In addition, M. D.  acknowledges the land on which the University of Toronto operates: For thousands of years it has been the traditional land of the Huron-Wendat, the Seneca, and most recently, the Mississaugas of the Credit River. Finally, S. C.  wishes to acknowledge that Queen's University is situated on traditional Anishinaabe and Haudenosaunee territory. We are grateful to be able to live, learn and play on these lands. We thank Ming Yang for correspondence and thoughts on the separation of RSGs from AGBs,
and Michael Irwin for providing us with the very nice UKIRT M33 data.  We are also grateful to our colleagues Georges Meynet, Sylvia Ekstr\"{o}m, and Cyril Georgy for allowing us to use their $z=0.006$ models in advance of publication.   We also appreciate correspondence with Bruce Draine concerning extinction maps of M31.  An anonymous referee made useful suggestions which improved this paper. P. M.'s work was partially supported by the National Science Foundation through AST-1612874. 
K. F. N.'s work was supported by a Cottrell Scholar Award from the Research Corporation for Scientific Advancement granted to E. M. L.  Both
M. R. D. and S. C.  acknowledge generous funding from the Natural Science and Engineering Research Council of Canada via the Discovery Grant program.
M. R. D. also acknowledges support from the Canada Research Chairs Program, the Canadian Institute for Advanced Research, and the Dunlap Institute at the University of Toronto.
This publication makes use of data products from the Two Micron All Sky Survey, which is a joint project of the University of Massachusetts and the Infrared Processing and Analysis Center/California Institute of Technology, funded by the National Aeronautics and Space Administration and the NSF.  


\clearpage

\clearpage

\clearpage

\clearpage

\clearpage

\clearpage

\clearpage

\clearpage

\clearpage

\clearpage

\clearpage
\begin{deluxetable}{l  c c}
\tablecaption{\label{tab:Data} Characteristics of Imaging Data}
\tablewidth{0pt}
\tablehead{
&\colhead{M31}
& \colhead{M33} 
}
\startdata
Telescope & 3.6-m CFHT & 3.8-m UKIRT \\
Exp. time (s) \\
\multicolumn{1}{c}{$J$} & 752 & 150 \\
\multicolumn{1}{c}{$K_s$} & 650 & 270 \\
Scale (\arcsec/pixel) & 0.3 & 0.4 \\
Coverage (deg$^2$) & 5.0 & 3.0 \\
Seeing (\arcsec) & 0.5-0.6 & 0.7-1.1 \\
Bright limit\tablenotemark{a} \\
\multicolumn{1}{c}{$J$} & 14.3 & 12.0 \\
\multicolumn{1}{c}{$K_s$} & 13.3 & 12.0 \\
Number of stars &  712,716 &    121,328 \\
\enddata
\tablenotetext{a}{2MASS was used to supplement the list for brighter stars.}
\end{deluxetable}

 \clearpage
\begin{deluxetable}{l l r}
\tabletypesize{\scriptsize}
\tablecaption{\label{tab:FunFacts} Adopted and Derived Relations}
\tablewidth{0pt}
\tablehead{
&\colhead{Relation}
&\colhead{Source}
}
\startdata
\sidehead{Adopted Distance:}
\phantom{MakeSomeSpace}& M31: 760~kpc (DM=24.40~mag)& 1 \\
 & M33: 830~kpc  (DM=24.60~mag)& 1 \\
\sidehead{Reddening Relations:}
&$A_K =0.12 A_V = 0.686 E(J-K)$ & 2 \\
&$E(J-K) = A_V/5.79 $ & 2 \\
\sidehead{RSG Photometric Criteria:}
\multicolumn{1}{r}{M31:}  
        &$15.5<K_s \leq 17.0$: $K_s \geq K_{s0}$ and $K_s \leq K_{s1}$ & 3 \\
        &$K_s\leq15.5$: $J-K_s\geq 0.984$ and $K_s \leq K_{s1}$ & 3 \\
                &$K_s\leq 14.4$ and $(J-K_s)\leq 1.9$: $J-K_s\geq 0.984$ & 3 \\
        &$K_{s0}=28.62-13.33(J-K_s)$  & 3,4\\
        &$K_{s1}=31.95-13.33(J-K_s)$  & 3,4\\
\multicolumn{1}{r}{M33:}
        &$15.0<K_s \leq 17.0$: $K_s \geq K_{s0}$ and $K_s \leq K_{s1}$ & 3 \\
        &$K_s\leq15.0$: $J-K_s\geq 0.963$ and $K_s \leq K_{s1}$ & 3 \\
                &$K_s\leq 14.6$ and $(J-K_s)\leq 1.9$: $J-K_s\geq 0.917$ & 3 \\
        &$K_{s0}=28.04-13.542(J-K_s)$  & 3,4\\
        &$K_{s1}=31.56-13.542(J-K_s)$  & 3,4\\
\sidehead{Adopted Extinction:}
        & $K_s>14.5$: $A_V=0.75$ & 3 \\
        & $K_s\leq 14.5$ and $K_s\leq K_{s1}$: $A_V=0.75-1.26\times(K_s-14.5)$ & 3 \\
        & $K_s\leq 14.5$ and $K_s\geq K_{s1}$: $A_V=0.75+5.79\times\Delta(J-K_s)$ & 3 \\
        & M31: $\Delta(J-K_s)=(J-K_s) - (30.29-K_s+0.686 (J-K_s))/14.02$ & 3 \\
        & M33: $\Delta(J-K_s)=(J-K_s) - (30.14-K_s+0.686 (J-K_s))/14.23$ & 3 \\
\sidehead{Conversion of 2MASS ($J, K_s$) to Standard System ($J, K$):}
&$K=K_s + 0.044$ & 5\\
&$J-K = (J-K_s+0.011)/0.972$ & 5\\
\sidehead{Conversion to Physical Properties (Valid for 3500-4500 K):}
&M31: $T_{\rm eff} = 5643.5 - 1807.1 (J-K)_0$ & 3\\
&M33: $T_{\rm eff} = 5606.6 - 1713.3 (J-K)_0$ & 3\\
&${\rm BC}_K = 5.495 - 0.73697 \times T_{\rm eff}/1000$ & 3\\
&$K_* = K - A_K$ & \nodata \\
&$M_{\rm bol} = K_* + {\rm BC}_K - DM$ & 1 \\
&$\log L/L_\odot =(M_{\rm bol}-4.75)/-2.5$ & \nodata \\
\enddata

\tablerefs{1--\citealt{vandenbergh2000}; 2--\citealt{Schlegel1998};
3--This paper; 4--\citealt{2006AA...448...77C}; 5--\citealt{Carpenter}
}
\end{deluxetable}

 \clearpage
 
\rotate
\begin{deluxetable}{l l l l l l l l l l l l l l l l }
\tabletypesize{\scriptsize}
\tablecaption{\label{tab:M31RSGs} Photometrically Selected RSGs in M31}
\tablewidth{0pt}
\tablehead{
\colhead{M31RSG\#}
& \colhead{$\alpha_{2000}$}
& \colhead{$\delta_{2000}$}
& \colhead{$\rho$\tablenotemark{a}}
& \colhead{$K_s$}
& \colhead{$\sigma_{K_s}$}
& \colhead{$J-K_s$}
& \colhead{$\sigma_{J-K_s}$}
& \colhead{\#obs}
& \colhead{Gaia\tablenotemark{b}}
& \colhead{$A_V$}
& \colhead{Teff\tablenotemark{c}}
& \colhead{$\log L/L_\odot$\tablenotemark{d}}
& \multicolumn{3}{c}{LGGS} \\ \cline{13-15}
&&&&&&&&&&&&\colhead{ID} & \colhead{$V$} & \colhead{Type}
}
\startdata
6832 & 00 40 50.843 &+41 06 48.30 &0.61& 17.368 & 0.038 & 1.080 & 0.052  &1 &2 & 0.75 & 3850 &3.67 &\nodata &\nodata &\nodata\\
6833 & 00 40 50.845 &+41 45 49.99 &1.61& 15.420 & 0.012 & 1.197 & 0.030 & 2 &2 & 0.75 & 3650 &4.38 &\nodata &\nodata &\nodata\\
6834 & 00 40 50.846 &+41 05 40.98 &0.58 &14.793 & 0.004 & 1.031&  0.006  &1& 0  &0.75  &3950 &4.72 &J004050.87+410541.1  &19.72 &RSG\\
6835 & 00 40 50.865 &+40 39 46.27 &0.47& 17.260 & 0.034&  1.093 & 0.048  &1 &2  &0.75  &3850 &3.70 &J004050.84+403946.3 & 22.07&\nodata\\
6836 & 00 40 50.874 &+40 46 21.81& 0.39 &17.260 & 0.037 & 1.077 & 0.148  &1 &2  &0.75  &3850 &3.71 &J004050.87+404621.6  &22.59&\nodata\\
\enddata
\tablecomments{Table 3 is published in its entirety in the machine-readable format. A portion is shown here for guidance regarding its form and content.}
\tablenotetext{a}{Galactocentric distance.  Assumes Holmberg radius of 95.3 arcmin, inclination 77.0 deg, and a position angle of the major axis of 35.0 deg.  At a distance of 760 kpc, a $\rho$ of 1.00 corresponds to 21.07 kpc.}
\tablenotetext{b}{Gaia flag: 0=probable member, 1=uncertain membership; 2=no Gaia data.}
\tablenotetext{c}{Typical uncertainty 150~K.}
\tablenotetext{d}{Typical uncertainity 0.05~dex.}
\end{deluxetable}

\clearpage
\rotate
\begin{deluxetable}{l l l l l l l l l l l l l l l l }
\tabletypesize{\scriptsize}
\tablecaption{\label{tab:M33RSGs} Photometrically Selected RSGs in M33}
\tablewidth{0pt}
\tablehead{
\colhead{M33RSG\#}
&\colhead{$\alpha_{2000}$}
& \colhead{$\delta_{2000}$}
& \colhead{$\rho$\tablenotemark{a}}
& \colhead{$K_s$}
& \colhead{$\sigma_{K_s}$}
& \colhead{$J-K_s$}
& \colhead{$\sigma_{J-K_s}$}
& \colhead{\#obs}
& \colhead{Gaia\tablenotemark{b}}
& \colhead{$A_V$}
& \colhead{Teff\tablenotemark{c}}
& \colhead{$\log L/L_\odot$\tablenotemark{d}}
& \multicolumn{3}{c}{LGGS} \\ \cline{13-15}
&&&&&&&&&&&&\colhead{ID} & \colhead{$V$} & \colhead{Type}
}
\startdata
854&01 32 19.96 & +30 37 18.6 & 1.04 &17.114 & 0.024&  0.981  &0.038  &4& 2 & 0.75 & 4100 &3.92 &\nodata &\nodata &\nodata \\
855&01 32 20.03 & +30 34 18.0 & 1.01 &16.874 & 0.019&  0.922 & 0.030  &4 &0 & 0.75  &4200 &4.04 &J013220.05+303418.1 & 20.76&\nodata \\
856&01 32 20.32  &+30 29 37.9 & 0.98 &16.229 & 0.011&  0.924 & 0.018  &4 &0 & 0.75 & 4200 &4.30 &J013220.36+302938.0  &19.69 &RSG\\
857&01 32 20.69 & +31 21 52.2 & 2.23 &16.124&  0.013&  0.884 & 0.019  &2 &0 & 0.75  &4250 &4.36 &\nodata &\nodata &\nodata \\
858&01 32 20.95 & +30 24 16.2&  0.98 &17.397&  0.031&  0.856 & 0.051  &4& 2 & 0.75 & 4300 &3.87&\nodata &\nodata &\nodata \\
\enddata
\tablecomments{Table 4 is published in its entirety in the machine-readable format. A portion is shown here for guidance regarding its form and content.}
\tablenotetext{a}{Galactocentric distance.  Assumes Holmberg radius of 30.8 arcmin, inclination 56.0 deg, and a position angle of the major axis of 23.0 deg.  At a distance of 830 kpc, a $\rho$ of 1.00 corresponds to 7.44 kpc.}
\tablenotetext{b}{Gaia flag: 0=probable member, 1=uncertain membership; 2=no Gaia data.}
\tablenotetext{c}{Typical uncertainty 150~K.}
\tablenotetext{d}{Typical uncertainty 0.05~dex.}
\end{deluxetable}

\clearpage

	
\bibliographystyle{aasjournal}
\bibliography{masterbib}

\end{document}